%% file: countingNew.tex
\documentclass{lmcs}
\usepackage{amssymb,amsmath}
\usepackage{epsfig}

\usepackage{thm-restate}
\usepackage{hyperref}

\newtheorem{claim}[thm]{Claim}
\newtheorem{observation}[thm]{Observation}

\newcommand{\sconnected}[1]{\mbox{[$#1$]-}connected}
\newcommand{\adj}[1]{\mbox{[$#1$]-}adjacent}
\newcommand{\component}[1]{\mbox{[$#1$]-}component}

\newcommand{\connected}[1]{\mbox{[$#1$]-}connected}

\newcommand{\nodes}{\mathit{nodes}}
\newcommand{\edges}{\mathit{edges}}

\newcommand{\vertices}{{\mathit{vertices}}}
\newcommand{\A}{\mathcal{A}}
\newcommand{\B}{\mathcal{B}}
\newcommand{\HD}{{\rm HD}}
\newcommand{\HG}{{\mathcal H}}
\newcommand{\FH}{{\mathcal FH}}
\newcommand{\Frontiers}{\mathit{Frontiers}}
\newcommand{\JT}{J\!T}

\renewcommand{\root}{\mathit{root}}

\newcommand{\U}{\mathcal{U}}

\newcommand{\DB}{{\mathrm{D}}}

\newcommand{\onDB}{{\mbox{\rm \tiny D}}}

\newcommand{\atoms}{\mathit{atoms}}

\newcommand{\views}{\mathit{views}}

\newcommand{\cores}{{\tt cores}}

\newcommand{\V}{\mathcal{V}}
\newcommand{\VQ}{\V_Q}

\newcommand{\free}{\mathit{free}}
\newcommand{\vars}{\mathit{vars}}

\newcommand{\sharpcovered}{{\sf \#}-covered}
\newcommand{\sharpdecomposition}{{\sf \#}-decomposition}
\newcommand{\sharpHD}{{\sf \#}\mbox{-}hypertree\ decomposition}
\newcommand{\adorn}{\mathit{color}}

\newcommand{\countProblem}{\mathrm{count}}

\newcommand{\Fr}{\mathit{Fr}}
\newcommand{\boA}{\mathbf{C}}
\newcommand{\boADB}{\mathbf{C}_h}

\newcommand{\form}{\mathit{form}}

\newcommand{\kHD}{\mathcal{C}_k}
\newcommand{\kNFHD}{\mathcal{C}_k^{\it nf}}

\newcommand{\size}[1]{\parallel\!\! #1\!\!\parallel }
\newcommand{\Pol}{\mbox{\rm P}}
\newcommand{\NP}{\mbox{\rm NP}}

\newcommand{\LCFL}{\mbox{\rm LOGCFL}}

\newcommand{\FPT}{\mbox{\rm FPT}}

\newcommand{\C}{\mathcal{C}}

\newcommand{\tuple}[1]{\langle#1\rangle}
\newcommand{\nop}[1]{}

\newcommand{\longv}[1]{}

\newcommand{\N}{\mathbb{N}}
\newcommand{\prom}[1]{\mathsf{prom}\textup{-}#1}
\newcommand{\param}[1]{\mathsf{param}\textup{-}#1}
\newcommand{\paramprom}[1]{\param{\prom{#1}}}
\newcommand{\powfin}{\wp_{\mathsf{fin}}}
\newcommand{\lpr}{\langle}
\newcommand{\rpr}{\rangle}
\newcommand{\sCQ}{\mathrm{\#CQ}}
\newcommand{\id}{\mathrm{id}}
\newcommand{\fadorn}{\mathit{fullcolor}}
\newcommand{\calC}{\mathcal{C}}
\newcommand{\calG}{\mathcal{G}}
\newcommand{\calN}{\mathcal{N}}
\newcommand{\calQ}{\mathcal{Q}}
\newcommand{\pn}[1]{\textsc{#1}}
\newcommand{\clique}{\pn{Clique}}
\newcommand{\sclique}{\pn{\#Clique}}

\newcommand{\graph}{\mathit{graph}}

\newcommand{\swone}{\mbox{\rm \#W[1]}}
\newcommand{\relB}{\mathrm{B}}
\newcommand{\simple}{\mathit{simple}}
\newcommand{\smallp}{\ensuremath{p\textup{-}}}

\begin{document}

\title[Counting Solutions to Conjunctive Queries]{Counting Solutions to Conjunctive Queries: Structural and Hybrid Tractability}

\author[H.~Chen]{Hubie Chen}[a]
\author[G.~Greco]{Gianluigi Greco}[b]
\author[S.~Mengel]{Stefan Mengel}[c]
\author[F.~Scarcello]{Francesco Scarcello}[a]
\address{King's College London}
\email{hubie.chen@kcl.ac.uk}
\address{University of Calabria}
\email{\{gianluigi.greco,francesco.scarcello\}@unical.it}
\address{Univ.~Artois, CNRS, Centre de Recherche en Informatique de Lens (CRIL)}
\email{mengel@cril-lab.fr}

\begin{abstract}
Counting the number of answers to conjunctive queries is a fundamental problem in databases that, under standard assumptions, does not have an efficient solution. 
The issue is inherently $\rm \#P$-hard, extending even to classes of \emph{acyclic} instances.

To address this, we pinpoint tractable classes by examining the structural properties of instances and introducing the novel concept of \emph{{\sf \#}-hypertree decomposition}. We establish the feasibility of counting answers in polynomial time for classes of queries featuring bounded {\sf \#}-hypertree width. Additionally, employing novel techniques from the realm of fixed-parameter computational complexity, we prove that, for bounded arity queries, the bounded {\sf \#}-hypertree width property precisely delineates the frontier of tractability for the counting problem. 
This result closes an important gap in our understanding of the complexity of such a basic problem for conjunctive queries and, equivalently, for constraint satisfaction problems (CSPs).

Drawing upon {\sf \#}-hypertree decompositions, a ``hybrid'' decomposition method emerges. This approach leverages both the structural characteristics of the query and properties intrinsic to the input database, including keys or other (weaker) degree constraints  that limit the permissible combinations of values. Intuitively, these features may introduce distinct structural properties that elude identification through the ``worst-possible database'' perspective inherent in purely structural methods.
\end{abstract}

\maketitle



\section{Introduction}

\subsection{Counting Solutions}

In this work, we delve into the complexity and propose algorithms for the problem $\sCQ$ of counting the number of answers to a conjunctive query (CQ) or, equivalently, the number of solutions to a constraint satisfaction problem (CSP), with respect to a given set of output variables. This problem is fundamental for the evaluation, optimization, and visualisation of database queries, but also
 for the applications of CSPs.
 
The problem is $\rm \#P$-hard~\cite{PS13}.  
Thus, our main goal is to identify those classes of instances whose structural properties allow solving the problem in polynomial time. Both the query (or the set of constraints to be satisfied) and the database are part of the input and must be taken into consideration in the complexity evaluation. 

In this section, we provide a simple and not overly formal introduction to the problem and the main contributions, giving us a roadmap for the subsequent, more technical sections.

\begin{figure*}[t]
 \centering
    \includegraphics[width=0.6\textwidth]{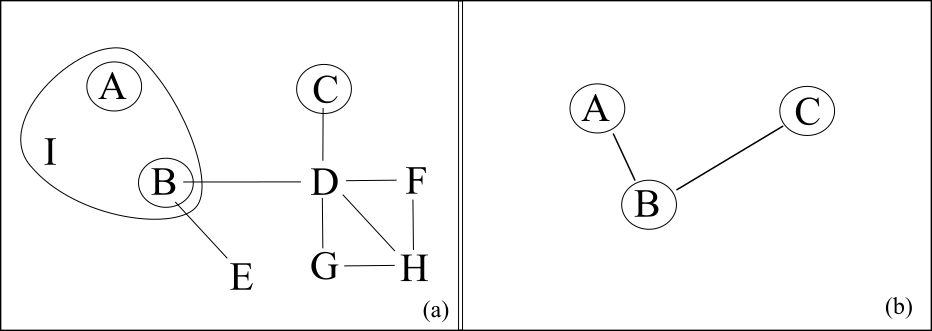}
  \caption{(a) Hypergraph $\HG_{Q_0}$ from Example~\ref{ex:esempio1}; and (b) its Frontier Hypergraph
   $\FH(Q_0,\{A,B,C\})$.}\label{fig:esempio1}
\end{figure*}

\begin{exa}\label{ex:esempio1}
Consider a database whose relations have the form:
\begin{itemize}
\item $mw(machine, worker\_id, machine\_hours)$: information on assignments machines to workers for a certain amount of time; 
\item $wi(worker\_id, worker\_info)$: information on workers;
\item $wt(worker\_id, task)$: assignments of tasks to workers;
\item $pt(project, task)$: tasks to be performed for each project;
\item $st(task, subtask)$: tasks and subtasks;
\item $rr(task, resource)$: tasks requirements.
\end{itemize}

We aim to count all sets of triples in the form $\tuple{A, B, C}$, where $A$ represents a machine assigned to a worker identified by the number $B$; and this worker is involved in tasks required by project $C$, which must satisfy additional conditions formally defined through the following conjunctive query $Q_{0}$:
$$
\hspace{-1mm}\begin{array}{ll}
   \exists D,E,F,G,H,I\  & mw(A,B,I)\wedge wt(B,D) \wedge wi(B,E) \wedge pt(C,D)\ \wedge\\
        &  st(D,F) \wedge st(D,G)\wedge rr(G,H) \wedge rr(F,H) \wedge rr(D,H).\\
\end{array}
$$

Figure~\ref{fig:esempio1} shows the hypergraph associated with ${Q_0}$.
The free variables of the query are interpreted (and called hereafter) output variables; for ${Q_0}$ they are $\{A\}$, $\{B\}$, and $\{C\}$ and are identified by circles in the figure. Hyperedges of cardinality 2 are depicted as lines.
 \hfill $\lhd$
\end{exa}

The objective is to efficiently determine the count of solutions without the need for explicit generation, considering that their number is typically exponential. It is crucial to emphasize that our interest lies solely in the free variables, in the example in quantifying machines, workers, and projects involved; we have no intention of considering all conceivable combinations of valid values for the remaining existential variables but require that one exists for the valid combinations of values assigned to the output variables.

The straightforward approach would involve computing the solutions and then isolating the variables of interest (free variables) for projection. However, this method incurs an exponential cost: even generating directly the restrictions of solutions to the variables of interest generally demands exponential time. Our aim, on the other hand, is to effectively tally only these solutions without their explicit generation while also preventing to count the same solution several times, a scenario that can arise due to different  combinations of the remaining existential variables.

\begin{figure*}[t]
 \centering
    \includegraphics[width=0.35\textwidth]{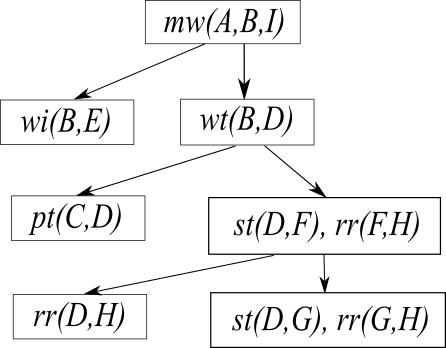}
  \caption{A width-2 hypertree decompositions of Hypergraph $\HG_{Q_0}$ in Example~\ref{ex:esempio1}.}\label{fig:HD1}
\end{figure*}

By looking at~Figure~\ref{fig:esempio1}, it is easy to see that our example query is cyclic, but its level of cyclicity is low. In~Figure~\ref{fig:HD1}, we present a hypertree decomposition of width 2 for the query. 
The notion of hypertree width, is based on organizing hyperedges (query atoms) into suitable clusters (subproblems) and arranging those clusters in a tree, called \emph{decomposition tree}. 
             This tree can be viewed as the {\em join tree} of an acyclic query equivalent to the original one, which is then evaluated via this tree, with a cost that is exponential in the cardinality of the largest cluster, also called {\em width} of the decomposition (it is $2$ in the example), and polynomial if the width is bounded by a constant (see, e.g., \cite{gott-etal-99,gott-etalGEMS}).
             In particular, it is possible to enumerate ``with polynomial delay'', only the solutions of interest (excluding the instantiations of existential variables)~\cite{GS13}. Unfortunately, this is insufficient for the counting problem, where we want to count the solutions without calculating them.       
             Pichler and Skritek~\cite{PS13} observed that classical decomposition methods are not helpful when projections are allowed, and showed that counting answers is~$\rm \#P$-hard, even for acyclic queries.
 In order to identify classes of tractable instances, it is necessary to consider the graphical structure of the output variables, in addition to the structure of the query hypergraph. 
 In fact, Durand and Mengel~\cite{DurandM15} showed that counting problems are tractable on classes of acyclic queries (or having bounded generalized hypertree width) having also bounded \emph{quantified star size}, a novel structural parameter which they introduced to measure how the variables that are not required in the output are spread in the query.

\subsection{Contributions}

\subsubsection*{$\#$-Hypertree decompositions and Tractable Queries}

We seek a novel type of structural decomposition that takes output variables into account. 

A straightforward approach might involve mandating that decompositions include a vertex where all output variables coexist. This would be tantamount to decomposing a hypergraph with an additional clique comprised of these variables. However, this would lead to decompositions with unnecessarily high width and thus to inefficient counting of query answers.

Instead, our approach is to pinpoint the smallest set of ``hidden'' relationships among output variables. These relationships are induced by how output variables interact with the existential variables within the given query. We define what we call the {\em Frontier Hypergraph} of existential variables, where the hyperedges precisely correspond to such relationships.

To understand the concept, let us consider the hypergraph depicted in Figure~\ref{fig:esempio1} (a) and imagine removing the free variables $\{A, B, C\}$.
Notice that this splits the hypergraph into three distinct connected components: one containing only variable $I$, one containing variable $E$, and one containing the set $\{D, F, G, H\}$. For each of these variables, we define its {\em frontier} as the set of free variables directly linked to the component in which it appears. Consequently, the frontier of $I$ is the set $\{A, B\}$, the frontier of $E$ is $\{B\}$, and the frontier of any existential variable in $\{D, F, G, H\}$ is $\{B, C\}$. The frontier hypergraph precisely consists of these frontiers as its hyperedges.
In Figure~\ref{fig:esempio1}(b), we show the frontier hypergraph for the query $Q_{0}$, labeled as $\FH(Q_0,\{A,B,C\})$. 
Note in particular that the variables $B$ and $C$, which are not directly linked in the hypergraph $\HG_{Q_{0}}$, are transitively connected through existential variables. In essence, if we wish to disregard the existential variables in the bottom-right part of the hypergraph, we should be able to identify the correct pairs of workers and projects. 
 However, there is no need to connect $A$ and $C$ since these two variables aren't part of a frontier for the existential variables. In fact, $I$ is solely related to $A$ and $B$, while the other variables are only connected to $B$ and $C$.

The initial idea, therefore, is to define a new notion of hypertree decomposition that considers both the ``base'' query hypergraph and this ``frontier'' hypergraph.

\begin{figure*}[t]
 \centering
    \includegraphics[width=0.6\textwidth]{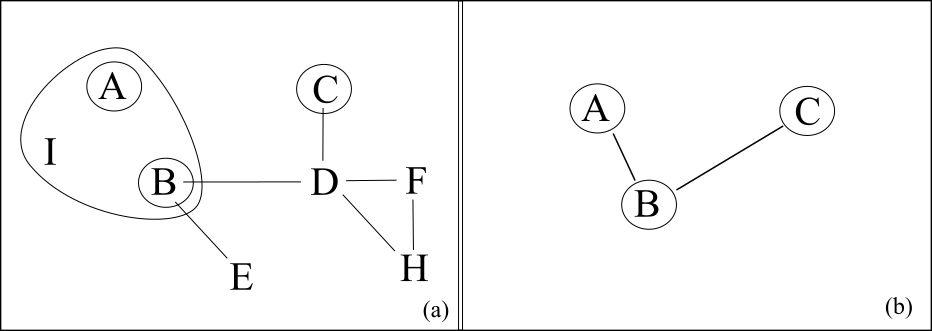} 
    \includegraphics[width=0.3\textwidth]{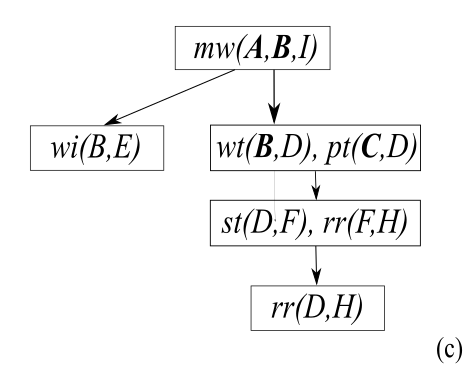}
  \caption{(a) Hypergraph $\HG_{Q_0'}$; (b) its Frontier Hypergraph
   $\FH(Q_0',\{A,B,C\})$; and (c) a width-2 $\#$-hypertree decomposition of the query in Example~\ref{ex:esempio1}.}
   \label{fig:HD2}
\end{figure*}

Another ingredient for identifying large tractable classes is query simplification: sometimes queries are equivalent to one of their subqueries. A smallest possible (homomorphically equivalent) subquery is called a {\em core}.
For instance, Figure~\ref{fig:HD2} (a) shows the hypergraph associated with a core of our example query, which does not include the hyperedges $\{D,G\}$ and $\{G,H\}$, and where variable $G$ does not occur at all. 
Intuitively, the subquery $st(D,G)\wedge rr(G,H)$ adds nothing to the query, as its role is already played by the ``equivalent''
 piece  $st(D,F)\wedge rr(F,H)$. 
An additional crucial point here is that we must not miss output variables, nor any of the pieces of the query that are ``relevant'' to these variables. For a given CQ $Q$, we identify the relevant substructures by defining {\em coloured cores}, more precisely any core of a modified query $\adorn(Q)$ where we add a fresh unary relation symbol for each output variable (depicted as circles, that is, unary hyperedges, in the example figures).

Then, the new notion of structural decomposition reads as follows.
\begin{defi}\label{def:sharpHypertree}
A query $Q$ is said to have a width-$k$ $\#$-{\em hypertree decomposition} if there exists a width-$k$ generalized hypertree decomposition for both the hypergraph of a core of $\adorn(Q)$ and its frontier hypergraph.
\end{defi}

In Figure~\ref{fig:HD2}, we show the hypergraph $\HG_{Q_0'}$ associated with a core of $\adorn(Q_{0})$ and a $\#$-hypertree decomposition for $Q_{0}$, which handles both the hyperedges of $\HG_{Q_0'}$ and those of the frontier hypergraph. In particular, note that the root node contains both variables $A$ and $B$, so that it covers the hyperedge $\{A,B\}$ of $\FH(Q_0',\{A,B,C\})$, and its right child contains both variables $B$ and $C$, so that it covers the other hyperedge $\{B,C\}$ of $\FH(Q_0',\{A,B,C\})$. 

Given a class $\boA$ of queries, let $\sCQ[\boA]$ be the problem of computing the number of answers of
 $Q$ on $\DB$, where $\DB$ is any database and $Q$ is any query belonging to $\boA$.
  We say that $\boA$ has bounded {\sf \#}-\emph{hypertree width} if there exists some finite natural number $k$ such that every query $Q\in\boA$ has a width-$k$ {\sf \#}-hypertree decomposition.

Our main tractability result states that all these instances are tractable, even though both computing a generalized hypertree decomposition and computing a core of a query are well-known	 $\NP$-hard problems.

\begin{restatable}{thm}{thmtractability}\label{thm:fptGHTW}
For every class of queries $\boA$ having bounded {\sf \#}-hypertree width,
$\sCQ[\boA]$ can be solved in polynomial-time.
\end{restatable}

To obtain this result, we show that for queries within $\boA$, it is possible to compute an appropriately colored core in polynomial time. Once such a core is available, we illustrate how to calculate the desired result within the broader tree-projection framework, as described in the following sections. This way, our tractability result
can be extended immediately to other structural decomposition methods, 
in particular to {\em fractional hypertree decompositions}~\cite{GM06}.

As we will show, the above property precisely defines the frontier of tractability for structures with bounded arity, thus  extending Grohe's seminal work on Boolean Conjunctive queries~\cite{G07} to counting problems. However, similar to Grohe's theorem, Theorem~\ref{thm:fptGHTW} is a
``promise'' based result: we obtain a correct answer for any given query having a{\sf \#}-hypertree width of at most $k$, but verifying this latter property in polynomial time is not possible unless $\Pol = \NP$.

For a non-promise result, we direct the interested reader to~\cite{GrecoS14}, a preliminary version of this work in which we introduce a weaker (polynomial-time) variant of the concept of {\sf \#}-hypertree decomposition. This variant ensures counting in polynomial time but does not establish a necessary condition for tractability.

\subsubsection*{Tractability result in a more general framework}

Structural decomposition methods such as hypertree decomposition or tree decomposition rely on the fact that it is possible to efficiently construct an acyclic instance equivalent to the given one using a fixed number k of atoms/constraints in the case of hypertree decomposition and a fixed number k of variables in the case of tree decomposition 
(see, e.g., \cite{gott-etal-99,GM06,G07,GMS07,GS17,GrecoS17,flum-etal-02,M10}). 
In practice, various techniques can be applied to address specific subproblems within the given instance. It's also plausible that some subproblems are readily accessible due to prior computations or other factors. Greco and Scarcello~\cite{GS08} have illustrated how, by employing the concept of  {\em tree projection}, tractability results can be attained within a more extensive framework. In this broader context, we have access to a collection of views (or solutions to subproblems) that can be utilized to either compute or, in the context of this paper, simply count the number of query answers.

In this scenario, we are presented with the query $Q$, a set of views $\V$, and a database $\DB$. 
As a technical requirement, the relations of views in $\DB$ cannot be ``more restrictive'' than the query, because otherwise some solutions of $Q$ on $\DB$ could be missed. A database respecting this requirement is said to be {\em legal} with respect to $Q$ and $\V$. 
Note that views obtained from subqueries of $Q$, as those defined according to structural decomposition methods, always have this property. 

We say that one hypergraph, $\HG$, is ``covered'' by another hypergraph $\HG'$, denoted as $\HG \leq \HG'$, when every hyperedge in $\HG$ is a subset (not necessarily proper) of a hyperedge in $\HG'$. 
 A {\em tree projection} is an acyclic hypergraph $\HG_{a}$ such that $\HG \leq \HG_{a} \leq\HG'$. It is sometimes also referred to as a ``sandwich hypergraph." 
 Whenever such a hypergraph $\HG_{a}$ exists, we say that the pair of hypergraphs $(\HG, \HG')$ has a tree projection.
  
\begin{restatable}{defi}{defTPdecomp}\label{def:decomposition}
A \emph{\sharpdecomposition\ of $Q$ w.r.t.~$\V$} is a tree projection $\HG_a$ for $(\HG_{Q'},\HG_\V)$ that covers the frontier hypergraph $\FH(Q',\free(Q))$, where $Q'$ is some core of $\adorn(Q)$.
Therefore, $\HG_a$ is an acyclic hypergraph such that $\HG_{Q'}\leq \HG_a\leq \HG_\V$ and $\FH(Q',\free(Q))\leq \HG_a$.
%

Whenever a \sharpdecomposition\ exists, we say that $Q$ is \emph{\sharpcovered\ w.r.t.~$\V$}. \hfill $\Box$
\end{restatable}

 \begin{figure}[t]
 \centering
    \includegraphics[width=0.8\textwidth]{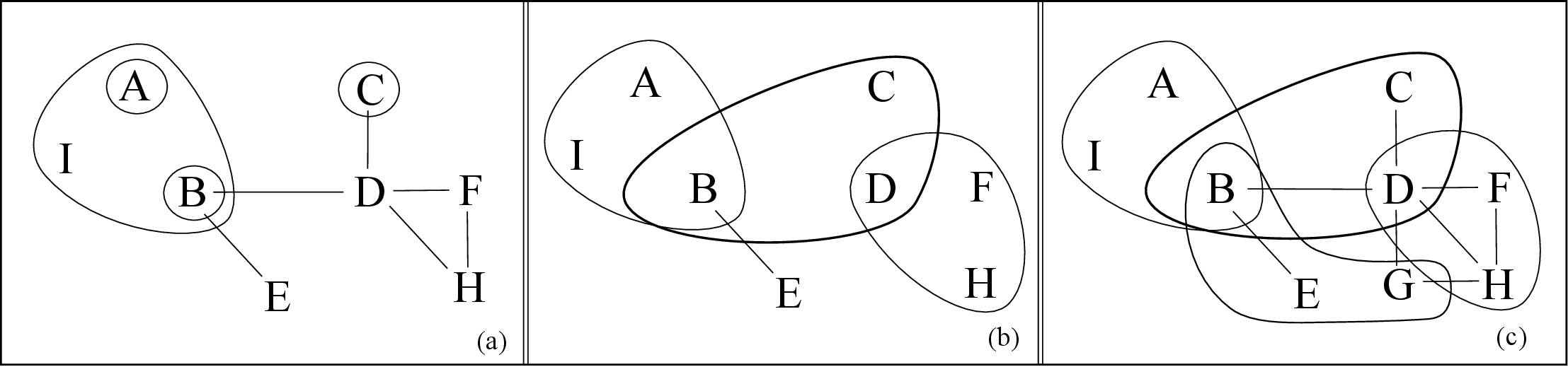}
  \caption{ (a)
  The hypergraph $\HG_{Q_0'}$; (b) A
  \sharpdecomposition\ of $Q_0$ w.r.t. $\V_0$; (c) The hypergraph $\HG_{\V_0}$.}\label{fig:esempio1tp}
\end{figure}
 
In Figure~\ref{fig:esempio1tp} (c), we show an example of views $\V_0$ (actually, their associated hypergraph),
i.e.~subproblems that we can address efficiently;
Figure~\ref{fig:esempio1tp} (c) shows a tree projection for the running example with respect to these views and the coloured core $Q_{0}'$ of $Q$. Note in particular that this hypergraphs covers both the hyperedges in $\HG_{Q_0'}$ and those belonging to the frontier hypergraphs (shown in Figure~\ref{fig:HD2} (b)).

 Intuitively,  a \sharpdecomposition\ identifies a way of exploiting the available tractable subproblems, such that
 both the relationships among all variables and the relationships among the output variables can be evaluated easily, as we do for acyclic instances.

It is worthwhile noting that, in this general context, the cores are not equivalent to each other. In fact, our instance has another symmetric core, in which $\{D, G\}$ and $\{G, H\}$ remain, and $\{D, F\}$ and $\{F, H\}$ disappear. However, this other core does not admit a tree projection with respect to the resources available in $\V_0$, because among these resources there is no view that is able to cover the triangle $\{D,G,H\}$.

In Section~\ref{sec:tractability-tp}, we show that the problem of {\em counting query answers is feasible in fixed-parameter polynomial time} (with the query size as fixed-parameter), or in polynomial time (no-promise) if a  \sharpdecomposition\  is additionally presented with the query instance.


\subsubsection*{Hybrid Decompositions}

In addition to structural features, it can sometimes be useful to leverage any characteristics of the databases being queried. For instance, there are typically keys on relations, which imply functional dependencies, or particularly selective attributes.

In these cases, it can be advantageous to consider some of the existentially quantified variables as if they were output variables. On one hand, this means that we should handle these ``pseudo output'' variables appropriately, incurring an overhead that depends on the values linking them to the actual free variables. On the other hand, larger sets of output variables can result in less intricate frontier hypergraphs, and therefore, fewer additional constraints compared to the original hypergraph.

Consider a query $Q$, with free variables $F$, over a database $\DB$.
For a relation $r$ and a set of variables $X$, the degree of a tuple $t$ is the number of ways to extend the values of the sub-tuple $t[X]$ to a complete tuple $t$ occurring in $r$; 
the degree $\deg_{\DB}(X,r)$ is the maximum degree over the tuples in $r$.
This notion is naturally extended to a hypertree $\HD=\tuple{T,\chi,\lambda}$:
For a vertex $v$ of $T$, consider the relation $r_{v}$ associated with $v$ according to the hypertree decomposition approach,
that is, $r_v=\ \pi_{\chi(v)}(\bowtie_{q\in \lambda(v)} q^\onDB)$. Consider its projection $\pi_{F} (r_{v})$ onto the variables in $F$. We say that a tuple $t$ in this projected relation has degree $b$ if it occurs $b$ times in the selection of $r_{v}$ w.r.t. $t$, that is, it occurs in $r_{v}$ with $b$ different extensions to the full set of variables occurring in $\chi(v)$. Define $\deg_{\DB}(F,v)$ to be the maximum degree over the tuples in $r_{v}$.

The maximum degree of the free variables over the vertices of $\HD$ is the \emph{boundness} of $\DB$ w.r.t.~$\HD$, denoted by
$\mathit{bound}(\DB,\HD)$. If we do not deal with the frontier hypergraph as shown above, we must keep information about the possible extensions of the free variables, which leads to a factor in the computation time that is exponential in ${\mathit{bound}(\DB,\HD)}$
(see Theorem~\ref{thm:mainHybrid}).

The idea is to leverage the knowledge of the input database to identify sets of variables with a very low degree. This way, we can assess whether certain existential variables might be more effectively treated as free variables.

\begin{figure*}[h]
 \centering
    \includegraphics[width=0.6\textwidth]{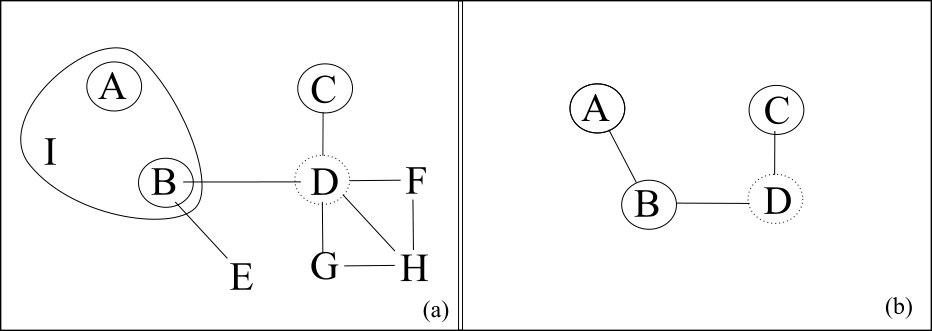}
  \caption{(a) Hypergraph $\HG_{Q_0}$ with the pseudo-free variable $D$; and (b) its Frontier Hypergraph
   $\FH(Q_0,\{A,B,C,D\})$.}\label{fig:esempio-hybrid-intro}
\end{figure*}

\begin{exa}\label{ex:hybrid-intro}
Consider again the instance in Example~\ref{ex:esempio1}, 
and assume that in relation $wt(worker\_id, task)$ each worker is involved in one task or in a few tasks, and that in relation 
$pt(project, task)$ each project comprises a few (main) tasks, while we have many subtasks and many resources involved in these projects.
That is, with reference to query $Q_{0}$, both degrees $\deg_{\DB}(B,wt)$ and $\deg_{\DB}(C,pt)$ are very small.

In this case, the variable $D$ associated with tasks can be seen as a free variable, while it is instead an existential variable (we are not interested in the number of tasks). In this modified scenario where $D$ is not existential, we get rid of its frontier (that would link  $B$ and $C$); in its new role, $D$ is the one variable in the frontier of the existential variables $F$, $G$, and $H$ (or just two of them if we consider cores of $Q_{0}$.  Figure~\ref{fig:esempio-hybrid-intro} shows the new frontier hypergraph. Note that all its edges are subsets of the original hyperedges, which means that every hypertree decomposition of the query hypergraph (as the one shown in Figure~\ref{fig:HD1}) automatically covers the frontier hypergraph. 
In this example, it implies that we don't have to explicitly calculate the pairs of workers and projects, in contrast to the hypertree decomposition in Figure~\ref{fig:HD2}. In fact, our intention is to count them without the need for explicit computation.
\end{exa}

In Section~\ref{sec:ht} we define the new hybrid notion of
  {\sf \#}$_b$-hypertree decomposition, which is based on the above idea to combine data information with the structural approach to deal with existential variables.
    Note that any candidate set of pseudo-free variables lead to different structural properties of the query, so that we have an additional source of complexity here.
Nevertheless, we show that dealing with this notion is fixed-parameter tractable, too.

\subsubsection*{Limits of Tractability in Counting Query Answers}

We also complement the algorithmic results discussed above with lower bounds. When considering classes of instances with bounded arity, we show that  one can count the query answers in polynomial time only for classes with bounded {\sf \#}-hypertree width.

The counting complexity of conjunctive queries without existential variables, where one seeks to count the valid assignments to all variables involved in the problem, was studied by Dalmau and Jonsson~\cite{DJ04}. They showed that bounded treewidth characterizes completely the tractable fragments of bounded arity for this problem.
As mentioned earlier, in counting problems (as well as in solution enumeration), it is crucial to focus solely on what happens to the variables of interest. The tractability frontier for bounded arity queries with existential variables has been refined by 
Durand and Mengel~\cite{DurandM15}, that characterizes the tractable classes for the problem restricted to \emph{simple} queries,
 where the same relation symbol cannot occur more than once in a query. We here solve the general case was left open in~\cite{DurandM15}.

The presence of quantified variables and atoms sharing the same relation symbol significantly complicates the problem, demanding additional proof techniques of independent interest.
To show our lower bound result, we use an adapted version of the case complexity framework introduced in \cite{Chen14}, tailored for studying the parameterized complexity of counting problems and for proving results independent of computability assumptions for the parameter. This framework is useful when investigating problems that encompass various potential restrictions on input instances, such as queries in this context.
 In particular, we introduce a suitable notion of reduction between case problems, referred to as {\em counting slice reduction}.
  By utilizing this framework, we can delineate the tractability frontier by drawing upon the hardness results presented 
   in~\cite{DJ04} and~\cite{DurandM15}.

Recall that a \emph{parameterized problem} is a set $J\subseteq \Sigma^*\times \mathbb{N}$. We say that $J$ is \emph{fixed-parameter tractable}
if there is a computable function $f: \mathbb{N}\to \mathbb{N}$, a constant $c\in \mathbb{N}$, and an algorithm that, given any pair
$(I,\pi)\subseteq \Sigma^*\times \mathbb{N}$
decides whether $(I,\pi)\in J$ in time $f(\pi)\times |I|^c$. The class of all fixed-parameter tractable problems is denoted by $\rm FPT$. There
is a hierarchy of parameterized complexity classes ${\rm FPT}\subseteq {\rm W[1]}\subseteq {\rm W[2]}\subseteq \cdots$. 
A complete problem for
${\rm W[1]}$ is the problem $\clique[\mathbb{N}]$ of deciding whether a graph $G$ has a clique of $k$ nodes, where $k$ is the parameter. This problem is not fixed-parameter tractable unless ${\rm FPT} = \rm W[1]$.

We show that the boundaries of tractability for the counting problem depict a trichotomy, in the spirit of the dichotomy stated in the landmark paper by~\cite{G07} about the complexity of answering Boolean conjunctive queries.
Let $\boA$ be a class of bounded arity queries. We denote by $\Frontiers(\boA)$ the frontier hypergraphs of these queries, that is,
  $\Frontiers(\boA) = \{ \FH(Q',\free(Q)) \mid  Q\in\boA $ and $Q'$ is some core of $\adorn(Q)\}$.
 
 The problem is in polynomial time (and hence also fixed-parameter tractable) if, and only if, $\boA$ has bounded {\sf \#}-hypertree width. If this is not the case but the class of frontier hypergraphs $\Frontiers(\boA)$ have bounded hypertree width, then
 the problem is {\rm W}[1]-equivalent, that is, there are parameterized reductions from and to the decision version of the parameterized Clique problem; otherwise, the problem is hard for the class of counting problems \#{\rm W}[1], as there exists a reduction from the parameterized problem of counting cliques.
 The formal statement, proved in Section~\ref{sec:frontier}, is as follows.
 
\begin{restatable}{thm}{thmtrichotomy} \label{thm:real-trichotomy}
Let $\boA$ be a computable class of bounded-arity queries.
\begin{enumerate}
 \item If the queries in $\boA$ have bounded {\sf \#}-hypertree width, then $\param{\sCQ[\boA]}$ is FPT.
 \item If the queries in $\boA$ have unbounded {\sf \#}-hypertree width, but the hypergraphs
  $\Frontiers(\boA)$  have bounded hypertree width, then
 $\param{\sCQ[\boA]}$ is equivalent to
 $\param{\clique[\mathbb{N}]}$ under counting FPT-reduction.

 \item Otherwise, there is a counting FPT-reduction from
$\param{\sclique[\mathbb{N}]}$ to
$\param{\sCQ[\boA]}$.
\end{enumerate}
\end{restatable}

The tractability frontier for the broader scenario involving classes with unbounded arities  is yet to be explored.
See the concluding remarks in Section~\ref{sec:conclusion} for a potential strategy to tackle this challenge.

\subsection{Related Work}

Since the publication of the conference versions and preliminary results presented in this paper~\cite{GrecoS14,ChenM15}, there have been several papers that build upon and extend these findings. One line of research extended our results to encompass unions of conjunctive queries~\cite{ChenM16,ChenM17},
 which poses additional problems since the same answer may appear in several parts of a union, so overcounting has to be avoided. 
Our trichotomy result has been recently refined in~\cite{DellRW19}, by considering two additional structural parameters called dominating star size and linked matching number: if the dominating star size is large, counting solutions is at least as hard as counting dominating sets, which yields a fine-grained complexity lower bound (under the Strong Exponential Time Hypothesis), as well as a \#{\rm W}[2]-hardness result in parameterized complexity; if the linked matching number is large then the problem is \#{\rm A}[2]-hard.

A noteworthy advancement in algorithms is the "Inside-Out" algorithm introduced in~\cite{FAQ-ngo}. 
This algorithm is designed for {\em functional aggregate queries} (FAQ) and can also be used to efficiently count solutions.
 Its runtime depends on the so-called {\rm FAQ}-width, which relies on an optimal variable ordering derived from fractional hypertree decompositions~\cite{GM06}. 
This research is further extended in~\cite{FAQ-subm}, where a relaxation of the set of polymatroids is introduced, giving rise to the counting version of Marx's submodular width~\cite{M10}, denoted as \#{\rm subw}. This new width concept falls between the submodular and fractional hypertree widths. Their new efficient FPT algorithm, named \#{\em PANDA}, achieves a runtime of 
  $\tilde O (N^{\#subw} )$ for evaluating FAQ over an arbitrary semiring. Note however that, in contrast to our results, the runtime is superpolynomial in the size of the query.
  
Our results have also been extended for the problem of counting query answers under integrity constraints and ontology-mediated queries~\cite{FeierLP21}. Berkholz et al.~study the data complexity of counting for (unions of) conjunctive queries under database updates~\cite{BerkholzKS17,BerkholzKS18}. We also mention the related problem of 
counting homomorphisms from hypergraphs of bounded generalised hypertree width, for which a logical characterisation is introduced  in~\cite{sche-23}.

A recent very exciting development has been the study of {\em approximate counting} for conjunctive queries. 
In this context, the emphasis is not on exact counts, but rather on extending the concept of tractability to encompass approximation schemes.
Quite recently, Arenas et al.~\cite{ArenasCJR22,ArenasCJR20} have shown that there exists a fully polynomial-time randomized approximation scheme (FPRAS) for every class of CQs with bounded hypertree width. In fact they also show that for queries defined over classes of graphs, bounded hypertree width is also a necessary condition for the existence of an FPRAS~\cite{ArenasCJR21}.
 This significant result was extended to classes of conjunctive queries with bounded fractional hypertree width in~\cite{FockeGRZ21}, where the authors conducted a comprehensive analysis of approximate counting for query answers, even when these queries include disequalities and negations.
These findings are of substantial interest as they, in a sense, add an additional piece to the puzzle presented in this paper: when the frontier hypergraph remains uncovered, exact solutions cannot be counted in polynomial time, but we can compute an approximation of this number, as long as the degree of cyclicity is low.

We would like to mention a related line of research dedicated to establishing precise complexity bounds for algorithms that count solutions for problems on graphs with bounded treewidth. For instance, in~\cite{Fock-etal23}, tight bounds are derived for a family of problems, including Independent Set, Dominating Set, Independent Dominating Set, and many others. Additionally, generalized coloring problems (list homomorphisms) are explored in~\cite{Fock-etal22}.


We also recall that  ``hybrid'' approaches (quite different from ours) have also been
proposed  for the related area of finite constraint satisfaction problems (CSPs), which are
in fact equivalent to conjunctive queries. 
In this context, tractable and intractable classes have been identified  by certain ``forbidden patterns'' in the constraint data (see, e.g., \cite{C13,CJ10} and the references therein).
These approaches identify islands of tractability in terms of properties expressed over the \emph{microstructure} of the instance,
which is a graph whose vertices represent the set of the possible assignments of values to variables, and whose edges are suitably induced from the original constraints.
Therefore, these methods are hybrid in that graph-theoretic notions are applied on a hybrid structure.
As working at the level of microstructure is hardly appropriate in database applications, we did not follow this approach. Rather, our approach in the last section is hybrid in that it exploits properties of the data to guide a decomposition with a subtle interplay between the two worlds. 
 Information regarding degree properties of databases has found successful application in query optimization~\cite{jogl-re-2018}.
 
 \subsection{Organization}

The remaining sections of the paper are structured as follows. Section~\ref{sec:framework} introduces some preliminary concepts related to conjunctive queries and structural decomposition methods. In Section~\ref{sec:structural}, we explore the notion of {{\sf \#}-covering} and its associated tractability outcomes within the broader context of {tree projection}. It is then refined in Section~\ref{sec:HDs}, specifically focusing on cases where generalized hypertree decompositions replace arbitrary tree projections, thereby introducing the concept of {\sf \#}-hypertree decomposition. The trichotomy result for the bounded arity case is established and examined in Section~\ref{sec:frontier}, while Section~\ref{sec:ht} defines and investigates the \emph{hybrid} notion of {\sf \#}$_b$-hypertree decompositions. Lastly, Section~\ref{sec:conclusion} provides some concluding remarks and outlines potential avenues for future research.
 
\section{Preliminaries}\label{sec:framework}

\noindent \textbf{Hypergraphs and Acyclicity.} A \emph{hypergraph} $\HG$ is a pair $(V,H)$, where $V$ is a finite set of nodes and $H$ is a set
of hyperedges such that, for each $h\in H$, $h\subseteq V$.
In the following, we denote $V$ and $H$ by $\nodes(\HG)$ and $\edges(\HG)$, respectively.

A hypergraph $\HG$ is {\em acyclic} (more precisely, $\alpha$-acyclic~\cite{fagi-83}) if, and only if, it has a join tree~\cite{bern-good-81},
i.e., a tree whose vertices are the hyperedges of $\HG$ such that, whenever a node $X\in V$ occurs in two hyperedges $h_1$ and $h_2$ of $\HG$,
then $h_1$ and $h_2$ are connected in $\JT$, and $X$ occurs in each vertex on the unique path linking $h_1$ and $h_2$. In words, the set of
vertices in which $X$ occurs induces a connected subtree of $\JT$.

%

\smallskip \noindent\textbf{Tree Projections.} For two hypergraphs $\HG_1$ and $\HG_2$, we say that $\HG_2$ \emph{covers} $\HG_1$, denoted by
$\HG_1\leq \HG_2$, if each hyperedge of $\HG_1$ is contained in at least one hyperedge of $\HG_2$. Assume that $\HG_1\leq \HG_2$. Then, a
\emph{tree projection} of $\HG_1$ with respect to $\HG_2$ is an acyclic hypergraph $\HG_a$ such that $\HG_1\leq \HG_a \leq \HG_2$. Whenever
such a hypergraph $\HG_a$ exists, we say that the pair of hypergraphs $(\HG_1,\HG_2)$ has a tree projection.

%

\smallskip \noindent\textbf{Relational Structures.}
Let $\U$ and $\mathcal{X}$ be disjoint infinite sets that we call the {\em universe of constants} and the \emph{universe of variables},
respectively. A (relational) vocabulary $\tau$ is a finite set of relation symbols of specified finite arities. A {\em relational structure}
$\A$ over $\tau$ (short: $\tau$-structure) consists of a universe $A\subseteq \U\cup \mathcal{X}$ and, for each relation symbol $r$ in $\tau$,
of a relation $r^\A\subseteq A^\rho$, where $\rho$ is the arity of $r$.

Let $\A$ and $\B$ be two $\tau$-structures with universes $A$ and $B$, respectively.
A {\em homomorphism} from $\A$ to $\B$ is a mapping $h: A \mapsto B$ such that $h(c)=c$ for each constant $c$ in $A\cap \U$, and such that, for
each relation symbol $r$ in $\tau$ and for each tuple $\tuple{a_1,\ldots,a_\rho}\in r^\A$, it holds that $\tuple{h(a_1),\ldots,h(a_\rho)}\in
r^\B$. For any mapping $h$ (not necessarily a homomorphism), $h(\tuple{a_1,\ldots,a_\rho})$ is used, as usual, as a shorthand for
$\tuple{h(a_1),\ldots,h(a_\rho)}$.

A $\tau$-structure $\A$ is a substructure of a $\tau$-structure $\B$ if $A \subseteq B$ and $r^\A \subseteq r^\B$, for each relation symbol $r$
in $\tau$.

\smallskip \noindent\textbf{Relational Databases.}
Let $\tau$ be a given vocabulary. A {\em database instance} (or, simply, a database) $\DB$ over $D\subseteq \U$ is a finite $\tau$-structure
$\DB$ whose universe is the set $D$ of constants. For each relation symbol $r$ in $\tau$, $r^\onDB$ is a \emph{relation instance} (or, simply,
relation) of $\DB$. Sometimes, we adopt the logical representation of a database~\cite{ullm-89,abit-etal-95}, where a tuple
$\tuple{a_1,...,a_\rho}$ of values from $D$ belonging to the $\rho$-ary relation (over symbol) $r$ is identified with the {\em ground atom}
$r(a_1,...,a_\rho)$. Accordingly, a database $\DB$ can be viewed as a set of ground atoms.

\smallskip\noindent\textbf{Conjunctive Queries.} A {\em conjunctive query} $Q$ is a first order formula
$\exists \bar {X} \Phi$ where $\Phi=r_1({\bf u_1})\wedge...\wedge r_m({\bf u_m})$ is a conjunction of atoms, $r_1,...,r_m$ (with $m > 0$) are
relation symbols, ${\bf u_1},...,{\bf u_m}$ are lists of terms (i.e., variables or constants), and $\bar {X}=X_1,...,X_n$ is the list of
quantified variables of~$\Phi$. We say that the query $Q$ is {\em simple} if every atom is defined over a distinct relation symbol. The
conjunction $\Phi$ is denoted by $\form(Q)$, while the sets of all atoms in $Q$ is denoted  by $\atoms(Q)$. For any set $A$ of atoms,
$\vars(A)$ is the set of all variables in $A$, and $\vars(Q)$ is used for short in place of $\vars(\atoms(Q))$. Moreover, we define $\free(Q)=$
$\vars(Q)\setminus$ $\{X_1,...,X_n\}$ as the set of the free variables in $Q$, which are understood as the output variables of the query.

For a database $\DB$ over $D$, $Q^\onDB$ denotes the set of all substitutions
$\theta:\vars(Q) \mapsto D$ such that for each $i\in\{1,...,m\}$,
it holds that $\theta'(r_{\alpha_i}({\bf u_i}))\in \DB$, where $\theta'(t)=\theta(t)$ if
$t\in \vars(Q)$ and $\theta'(t)=t$ otherwise (i.e., if the term $t$ is a constant).

Note that the conjunction $\form(Q)$ can be viewed as a relational structure $\mathcal{Q}$, whose vocabulary $\tau_Q$ and universe $U_Q$ are
the set of relation symbols and the set of terms occurring in its atoms, respectively. For each symbol $r_i\in\tau_Q$,  the relation
$r_i^{\mathcal{Q}}$ contains a tuple of terms ${\bf u}$
if and only if an atom of the form $r_i({\bf u})\in\atoms(Q)$
appears in the conjunction. In the special
case of simple queries, every relation $r_i^{\mathcal{Q}}$ of $\mathcal{Q}$ contains just one tuple of terms.
According to this view, elements in $Q^\onDB$ are in a one-to-one correspondence with homomorphisms from $\mathcal{Q}$ to $\DB_Q$, where the
latter is the (maximal) substructure of $\DB$ over the (sub)vocabulary $\tau_Q$. Hereinafter, for the sake of presentation, we freely use
interchangeably queries and databases with their relational structures, e.g., we may use $Q$ and $\DB$ in place of $\mathcal{Q}$ and $\DB_Q$.
Moreover, $||Q||$ and $||\DB||$ denote the sizes of the underlying relation structures, according to their standard encoding (see, e.g.,
\cite{G07}).

A query $Q'$ is a \emph{core} of $Q$ if it is a minimal substructure of $Q$ such that there is a homomorphism from $Q$ to $Q'$---note that the
quantification of the variables in $Q'$ is not taken into account here
(it will be accounted for by a certain form of coloring, to be introduced later).
The set of all cores of $Q$ is denoted by $\cores(Q)$. Elements in $\cores(Q)$ are
\emph{isomorphic} to each other.

\smallskip\noindent \textbf{Relational Algebra.}
For any set $W\subseteq \vars(Q)$ of variables and set $S$ of substitutions, we denote by $\pi_W(S)$ the set of the restrictions of the
substitutions in $S$ over the variables in $W$. If $\theta$ is a substitution with domain $W$, then we denote by $\sigma_{\theta}(S)$ the set
$\{\theta'\in S \mid \pi_W(\{\theta'\})=\{\theta\})$.
If $S_1$ and $S_2$ are sets of substitutions with domains $W_1$ and $W_2$, respectively, then we denote by $S_1 \bowtie S_2$ the set of all
substitutions $\theta$ over $W_1\cup W_2$ such that $\pi_{W_1}(\{\theta\})\subseteq S_1$ and $\pi_{W_2}(\{\theta\})\subseteq S_2$. We use $S_1
\ltimes S_2$ as a shorthand for $\pi_{W_1}(S_1\bowtie S_2)$.


\smallskip\noindent \textbf{Hypergraphs and atoms.} There is a very natural way to associate a hypergraph $\HG_\V=(N,H)$ with any set $\V$ of
atoms: the set $N$ of nodes consists of all variables occurring in $\V$; for each atom in $\V$, the set $H$ of hyperedges contains a hyperedge
including all its variables; and no other hyperedge is in $H$.
For a query $Q$, the hypergraph associated with $\atoms(Q)$ is denoted by $\HG_Q$.
%

\smallskip\noindent \textbf{Counting.}
According to the above notation, $\pi_{\free(Q)}(Q^\onDB)$ is the set of answers (over the
\emph{output} variables) of the conjunctive query $Q$ on the database $\DB$ .
We denote by $\countProblem(Q,\DB)$ the problem  of computing the cardinality of this set, i.e., the
number of (distinct) answers of $Q$ on $\DB$.

\section{Structural Analysis}\label{sec:structural}

Let $Q$ be a conjunctive query, and let $\V$ be a set of atoms, called {\em views}, each one defined over a specific relation symbol not occurring in $Q$. These
atoms play the role of additional resources that can be used to answer (counting problems on) $Q$, so that we are abstracting here all purely
\emph{structural decomposition methods} proposed in the literature, which just differ in how they define and build the set of such available
resources.

Following the framework in~\cite{GS17}, we say that $\V$ is a \emph{view set} for $Q$ if, for each atom $q\in\atoms(Q)$, $\V$ contains an atom
$w_q$ with the same list of variables as $q$ (but with a different relation symbol). Each atom in $\V$ is called a \emph{view}; in particular,
atoms of the form $w_q$ are called {\em query views}. The set of all query views will be denoted by $\views(Q)$.

Let $\DB$ be a database whose vocabulary includes all relation symbols in $Q$ and $\V$. In a structural decomposition method, query views are
initialized with the same tuples as their associated query atoms while, for any other view $w\in \V$, $w^\onDB$ is initialized by including
solutions of some subquery over the variables in $w$. Using such views for evaluating $Q$ is possible whenever $\DB$ is a \emph{legal} database
(on $\V$ w.r.t.~$Q$), i.e., (i)  $w_q^\onDB \subseteq q^\onDB$ holds, for each query view $w_q\in \views(Q)$; and (ii) $w^\onDB\supseteq
\pi_{\vars(w)}(Q^\onDB)$, for each view $w\in \V$. Intuitively, all original ``constraints'' are there, and views are not more restrictive than
the original query.

In this section, we look for ``structural'' restrictions to $Q$  guaranteeing that the problem $\countProblem(Q,\DB)$  can be efficiently
solved by exploiting the views in $\V$.

\subsection{\#-covered Queries}\label{sec:tp}

The first ingredient in the analysis is to ``color'' $Q$ in order to take into account in its structure the role played by the free variables.
Formally,  the {\em coloring of $Q$} is the query $\adorn(Q)$ having the same set of variables as $Q$ (with the same quantifications) and the
same atoms as $Q$, plus an additional atom $r_X(X)$ for each variable $X \in  \free(Q)$, with $r_X$ being a fresh relation symbol.
Thus, in a colored query, the atom $r_X(X)$ associated with a free variable $X$
allows us to distinguish its actual domain with respect to other variables occurring in query atoms over shared relation symbols.

\begin{figure}[h]
 \centering
    \includegraphics[width=0.99\textwidth]{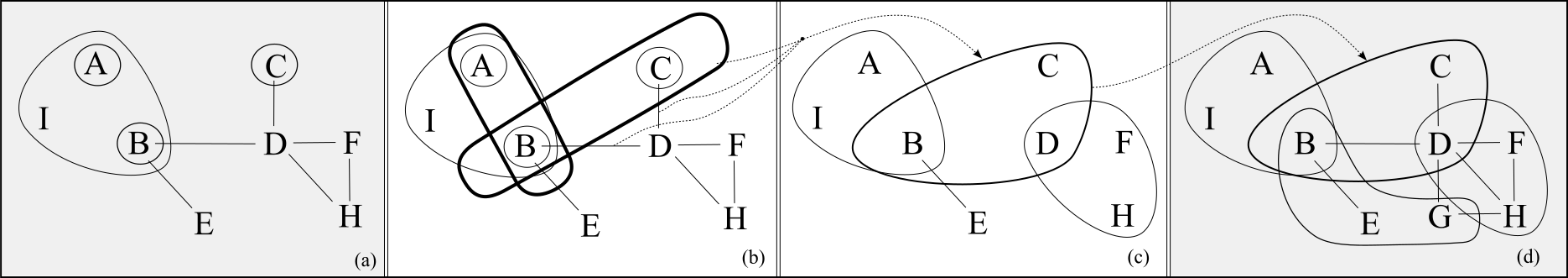}
  \caption{
  (a) The hypergraph $\HG_{Q_0'}$;   (b) $\HG_{Q_0'}$ plus the frontier hypergraph $\FH(Q_0',\free(Q_0'))$ (in bold);
    (c) A
  \sharpdecomposition\ of $Q_0$ w.r.t. $\V_0$; (d) The hypergraph $\HG_{\V_0}$.}\label{fig:esempio1b}\label{fig:esempioCovered}
\end{figure}

\begin{exa}\label{ex:esempio1b}
Consider again the conjunctive query $Q_0=\exists D,...,I\ \Phi$ with $\free(Q_0)=\{A,B,C\}$ in Example~\ref{ex:esempio1}, where
$$
\hspace{-1mm}\begin{array}{ll}
   \Phi = & mw(A,B,I)\wedge wt(B,D) \wedge wi(B,E) \wedge pt(C,D)\ \wedge\\
        &  st(D,F) \wedge st(D,G)\wedge rr(G,H) \wedge rr(F,H) \wedge rr(D,H).\\
\end{array}
$$

Then, $\adorn(Q_0)$ is the query $\exists D,...,I\ \Phi\wedge \Phi_f$, where
$$
\begin{array}{lll}
   \Phi_f & = & r_A(A)\wedge r_B(B) \wedge r_C(C).\\
\end{array}
$$

For the reader's convenience, we present in Figure~\ref{fig:esempio1b} the hypergraph associated with ${Q_0}$ previously depicted in Figure~\ref{fig:esempio1}, where the circled free variables $\{A\}$, $\{B\}$, and
$\{C\}$ are the singleton hyperedges additionally occurring in hypergraph $\HG_{\adorn(Q_0)}$.
 \hfill $\lhd$
\end{exa}

The concept of coloring is next applied within the tree projection framework. In particular, we focus on tree projections where a special
condition holds over the quantified variables, which keeps their interaction with free variables ``under control''. To state this condition,
some further concepts and notations are required.

Let $\HG$ be a hypergraph with $\nodes(\HG)\subseteq \vars(Q)$---so, we hereinafter use the terms \emph{nodes} and \emph{variables} interchangeably.
Let $X$ and $Y$ be two variables in $\vars(Q)$, and let $\bar W\subseteq \vars(Q)$ be a set of variables. We say that $X$ is \adj{\bar W}\ in
$\HG$ to $Y$ if there is a hyperedge $h\in \edges(\HG)$ such that $\{X,Y\}\subseteq (h\setminus \bar W)$. Moreover, $X$ and $Y$ are
\sconnected{\bar W}\ (in $\HG$) if there is a sequence $X=X_0,\ldots,X_\ell=Y$ such that $X_{i}$ is \adj{\bar W}\ to $X_{i+1}$, for each $i\in
\{0,...,\ell\mbox{-}1\}$.
Any maximal \connected{\bar W}\ non-empty set of variables  from $\nodes(\HG)\setminus \bar W$ is a \component{\bar W}.
For a set $H$ of hyperedges, let $\nodes(H)$ denote the set $\bigcup_{h\in H} h$, and for any \component{\bar W}\ $C$, let $\edges(C)$ denote the
set $\{ h\in \edges(\HG) \mid h\cap C\neq\emptyset\}$.

For any set $\bar W$ of variables (usually, the free variables), and for any variable $Y$, define the \emph{frontier} $\Fr(Y,\bar W,\HG)$ of $Y$ w.r.t.~$\bar W$ in $\HG$ (or,
shortly, $\Fr(Y,\bar W)$ if $\HG$ is understood) as follows: if $Y\in \bar W$, $\Fr(Y,\bar W,\HG)=\emptyset$; otherwise, $\Fr(Y,\bar W,\HG)=\bar W\cap
\nodes(\edges(C))$, where $C$ is the \component{\bar W} of $\HG$ where $Y$ occurs.
Note that all variables occurring in the same \component{\bar W} have the same frontier with respect to $W$.

\begin{figure*}[t]
 \centering
    \includegraphics[width=0.60\textwidth]{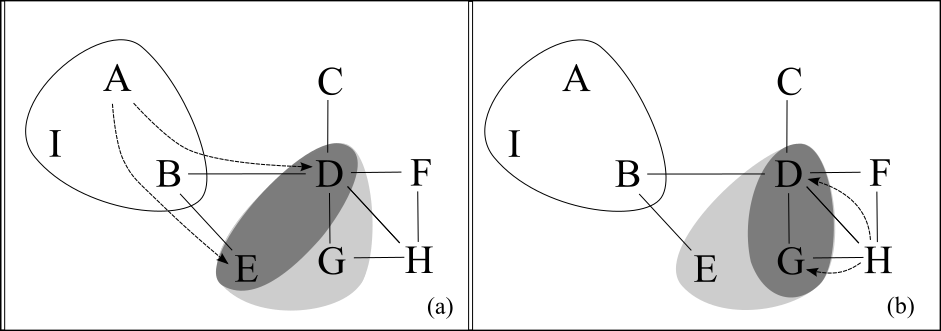}
  \caption{Structures in Example~\ref{ex:esempioFrontiersB}:  (a) The frontier $\Fr(A,\{D,E,G\},\HG_{Q_0})=\{D,E\}$; (b) The frontier $\Fr(H,\{D,E,G\},\HG_{Q_0})=\{D,G\}$.}\label{fig:esempio1f}
\end{figure*}

\begin{exa}\label{ex:esempioFrontiersB}
Consider again the hypergraph $\HG_{Q_0}$ in Example~\ref{ex:esempio1}. Then, Figure~\ref{fig:esempio1f}(a) shows the frontier $\Fr(A,\{D,E,G\},\HG_{Q_0})=\{D,E\}$.  Note, in particular, that $\{A,B,I\}$ is the \component{\{D,E,G\}} of $\HG_{Q_0}$ where $A$ occurs, and that $\edges(\{A,B,I\})$ consists of the hyperedges $\{A,B,I\}$, $\{B,D\}$, and $\{B,E\}$; hence, $\nodes(\edges(\{A,B,I\})=\{A,B,D,E,I\}$, so that $\Fr(A,\{D,E,G\},\HG_{Q_0})=\{A,B,D,E,I\}\cap \{D,E,G\} = \{D,E\}$. 
The frontier $\Fr(H,\{D,E,G\},\HG_{Q_0})= \{D,G\}$ is instead illustrated in Figure~\ref{fig:esempio1f}(b).
\hfill $\lhd$
\end{exa}

We next use the concept of frontier to define the frontier hypergraph of a query with respect to a set of free
variables.

\begin{defi}\label{def:frontierHypergraph}
Let $\bar W\subseteq \vars(Q)$ be a set of variables and $Q'$ be a query with $\vars(Q')\subseteq \vars(Q)$.
Then, the \emph{frontier hypergraph} $\FH(Q',\bar W)$ is the hypergraph $(\vars(Q')\cup \bar W, H)$, whose hyperedges are
 the frontiers of the variables of $Q'$ and the hyperedges of  $\HG_{Q'}$ covered by $\bar W$, that is, $H=\{ \Fr(Y,\bar W,\HG_{Q'}) \mid Y\in \vars(Q') \} \cup \{e \in \edges(\HG_{Q'}) \mid e \subseteq \bar W\}$.
\end{defi}

\begin{exa}\label{ex:esempioFrontiers}
Consider again the query $\adorn(Q_0)$ of Example~\ref{ex:esempio1} with free variables $\Lambda = \{A,B,C\}$, and the query $Q_0'=\Phi'\wedge \Phi_f$ where
$$
   \Phi' = mw(A,B,I)\wedge wt(B,D) \wedge wi(B,E) \wedge pt(C,D)
        \wedge  st(D,F) \wedge rr(F,H) \wedge rr(D,H).
   $$

The hypergraph $\HG_{Q'_0}$ is reported in Figure~\ref{fig:esempioCovered}(a). Note that $\HG_{Q'_0}$ coincides with $\HG_{\adorn(Q_0)}$, but for the hyperedges $\{D,G\}$ and $\{G,H\}$ whose corresponding relations are not in $Q_0'$; hence, $\vars(Q_0')=\vars(Q_0)\setminus\{G\}$.

Figure~\ref{fig:esempioCovered}(b) shows in bold the frontier hypergraph $\FH(Q_0',\free(Q_0'))$, where $\free(Q_0')=\free(Q_0)=\{A,B,C\}$, denoted by $\Lambda$. Its hyperedges are given by
hyperedges of $\FH(Q_0',\Lambda)$ are given by $\{ \Fr(Y,\Lambda,\HG_{Q'_0}) \mid Y\in \{A,B,C,D,E,F,H,I\} \} \cup \{e \in \edges(\HG_{Q'_0}) \mid e \subseteq \Lambda\}$.
In particular, note that $\Fr(A,\Lambda)=\Fr(B,\Lambda)=\Fr(C,\Lambda)=\emptyset$,
$\Fr(E,\Lambda)=\{B\}$, $\Fr(D,\Lambda)=\Fr(F,\Lambda)=\Fr(H,\Lambda)$=$\{B,C\}$, and $\Fr(I,\Lambda)$ $=$ $\{A,B\}$ hold. Moreover, $\{A\}$, $\{B\}$ and $\{C\}$ are the hyperedges of $\HG_{Q'_0}$ covered by $\Lambda$.~\hfill~$\lhd$
\end{exa}

Recall from the introduction the definition of \sharpdecomposition.
\defTPdecomp*

%
%

Note that, in the above definition, we talk about ``some'' core of the colored query rather than just about ``the'' core, because in this
general framework different cores might behave differently w.r.t.~the available views~\cite{GS17}.

\begin{exa}\label{ex:esempio2}
Consider again the query $\adorn(Q_0)$ of Example~\ref{ex:esempio1b}, and the query $Q_0'$ introduced in Example~\ref{ex:esempioFrontiers}. Note that $Q_0'$ is, indeed, a core of $\adorn(Q_0)$.

Then, consider a set $\V_0$ of views such that $\HG_{\V_0}$ is the hypergraph depicted in Figure~\ref{fig:esempio1b}(d), and check that the hypergraph
in Figure~\ref{fig:esempio1b}(c) is a tree projection of $\HG_{Q'_0}$ w.r.t.~$\HG_{\V_0}$.
For instance, note that the clique over $D$, $F$, and
$H$ is absorbed by a hyperedge in $\edges(\HG_{\V_0})$. 

In fact, the given tree projection covers the frontier hypergraph $\FH(Q_0',\free(Q_0))$, which has been depicted in Figure~\ref{fig:esempioCovered}(b). Indeed, in particular, note that the hyperedge $\{B,C,D\}$ covers not only the hyperedges $\{B,D\}$ and $\{C,D\}$, but also the hyperedge $\{B,C\}$ occurring in $\FH(Q_0',\free(Q_0))$. Therefore, the existence of this tree projection witnesses that $Q_0$ is {\sharpcovered\ w.r.t.~$\V_0$}.%
~\hfill~$\lhd$
\end{exa}

Computing a \sharpdecomposition\ is clearly $\NP$-hard in general, because both dealing with cores and tree projections are hard tasks.
However, in database applications it is important to study what happens if queries are not too large, which is often modeled by assuming that
they are fixed, as in data-complexity analysis. With this respect, we next provide a positive result for \sharpdecomposition s, by showing that
their computation, parameterized by the size of the query, is fixed-parameter tractable.

\begin{thm}\label{thm:fpt}
The following problem, parameterized by the query size, is in ${\rm FPT}$: Given a query $Q$ and a view set $\V$, decide whether there exists a
\sharpdecomposition\ of $Q$ w.r.t.~$\V$, and compute one (if any).
\end{thm}
\begin{proof}
Let the parameter $k=\size{Q}$, and let $Q'$ be any core of $\adorn(Q)$. Consider the hypergraph $\HG'=(V,H)$ where $V=\vars(Q')$ and
$H= \edges(\HG_{Q'})\cup \edges(\FH(Q',\free(Q)))$.
Clearly, covering both $\HG_{Q'}$ and the frontier hypergraph is equivalent to cover $\HG'$, so that the problem of computing a
 \sharpdecomposition\ of $Q$ w.r.t.~$\V$ is equivalent to the problem of computing
 a tree projection of $\HG'$ with respect to $\HG_\V$, that is, computing
an acyclic hypergraph $\HG_{a}$ such that $\HG'\leq \HG_a \leq \HG_\V$.
 This is known to be feasible in {\rm FPT}~\cite{GS17}, with the parameter being the number of nodes occurring in the smaller hypergraph, that is,
  $|\vars(Q')| \leq k$.

 It remains to find a right core $Q'$ to be used for the above computation; to this end, observe that the cores
of $\adorn(Q)$ can be enumerated in time $f({k}) O(n^{c})$, where $f(\cdot)$ is an exponential function and $c$ is a fixed constant.
\end{proof}

\subsection{Tractability of Counting}\label{sec:tractability-tp}

We can now relate the concept of {\sf \#}-covering with the tractability of the counting problem.
The computational costs here and elsewhere in the paper are reported assuming {unit cost for arithmetic operations}, as usual in this context.

\begin{thm}\label{thm:general}
Let $Q$ be a query, let $\V$ be a view set for it, and let $\DB$ be a legal database.
There exists a polynomial-time algorithm (w.r.t. $||Q||$, $||\DB||$, and $||\HG_a||$) that solves $\countProblem(Q,\DB)$, assuming that
 a \sharpdecomposition\ $\HG_a$ of $Q$ w.r.t.~$\V$ is presented
 along with each instance.
 \end{thm}
\begin{proof}
Let $Q_c$ be a core of $\adorn(Q)$, and let $Q'$ be its uncolored version where the special coloring atoms have been removed. Note that $Q'$ is
a subquery of $Q$ containing all its free variables (because of colors) and, as pointed out in~\cite{GS13}, $\pi_{\free(Q)}(Q'^\onDB) =
\pi_{\free(Q)} (Q^\onDB)$. Therefore, the solution of $\countProblem(Q',\DB)$ is the same as the solution of $\countProblem(Q,\DB)$. Moreover,
note that $Q_c$ and $Q'$ have the same hypergraph $\HG_{Q'}$ (but for the singleton edges associated with colors, which are subsets of other
edges, and hence are irrelevant as far as the existence of tree projections is concerned).

Let $\HG_a$ be a tree projection of $\HG_{Q'}$ w.r.t.~$\HG_\V$ such that $\HG_a$ covers the frontier hypergraph $\FH(Q',\free(Q))$. Let $\bar
C=\{C_1,...,C_e\}$ be the set of all the \component{\free(Q)}s of $\HG_{Q'}$, which form a partition of the quantified variables in
$\vars(Q)\setminus\free(Q)$. The property that all frontiers are covered entails that, for every $C_i\in\bar C$, there is a hyperedge $h_{C_i}$
of $\HG_a$ such that $h_{C_i}\supseteq \Fr(Y,\free(Q),\HG_{Q'})$, for any $Y\in C_i$. Indeed, note that
$\Fr(Y_s,\free(Q),\HG_{Q'})=\Fr(Y_j,\free(Q),\HG_{Q'})$ for each pair $Y_s,Y_j\in C_i$, so that the frontier is unique for every connected
component. Without loss of generality, we may assume that $h_{C_i}=\Fr(Y,\free(Q),\HG_{Q'})$, otherwise we could add such a hyperedge to
$\HG_a$ obtaining a new tree projection with the desired property. By definition of frontier, $h_{C_i}\subseteq \free(Q)$ (it contains no
quantified variables).

Define now the query $Q_a$ associated with the tree projection $\HG_a$ as the simple acyclic query having precisely one atom
$\mathit{atom}_j(\vec{h_j})$ for each hyperedge $h_j$ of $\HG_a$, where $\mathit{atom}_j$ is a fresh relation symbol and $\vec{h_j}$ is a list
with the variables occurring in $h_j$. It is well-known that, given any legal database $\DB$ for $\V$ and $Q$ (and hence for $Q'$),  we can
compute in polynomial time a database $\DB_a$ for $Q_a$ such that: $Q_a$ is equivalent to $Q'$, that is, $Q_a^{\onDB_a}=Q'^{\onDB}$; and
$\DB_a$ is pairwise consistent and hence global consistent, because $Q_a$ is acyclic. In particular, this can be done by enforcing pairwise
consistency (through semijoin operations) over all pairs of views in $\V$ until a fixpoint is reached. Let $\DB_\ell$ be the resulting database
for the views in $\V$. As shown in~\cite{GS17}, after this operation, whenever there exists a set of variables $\bar O$ such that $\bar
O\subset h_{\bar O}$ for some $h_{\bar O}\in\edges(H_a)$ (these sets are called tp-covered in ~\cite{GS17}), we know that, for every view
$w\in\V$ such that $\bar O\subseteq \vars(w)$, $\pi_{\bar O} (w^{\onDB_\ell})= \pi_{\bar O} (Q'^{\onDB})$.
Thus, we can define the database $\DB_a$ for $Q_a$ in such a way that its relations are suitable ``projections'' of view relations in the
pairwise consistent database $\DB_\ell$. More precisely, for every atom $\mathit{p}_j$ in $Q_a$, we set $\mathit{p}_j^{\onDB_a}= \pi_{h_j}
(w_j^{\onDB_\ell})$, where $h_j = \vars(\mathit{p}_j)$ and $w_j$ is any of its covering relation in $\V$, so that $h_j\subseteq \vars(w_j)$
holds. By the above property of the database $\DB_\ell$, note that every atom relation occurring in $Q_a$ contains all and only those tuples
that occur in query answers, formally, $\mathit{atom}_j(\vec{h_j})^{\onDB_a}= \pi_{h_j} (Q'^{\onDB})$. Therefore, adding any of these atoms to
$Q'$ does not change the query. In particular we would like to add all the atoms associated with the components of quantified variables in
$\bar C$, so that we get a new query $Q''= Q' \bigwedge_{C_i\in\bar C} \mathit{atom}_{C_i}(\vec{h_{C_i}})$. Accordingly, let $\DB''$ be the
database for $Q''$ obtained by adding to $\DB$ the relations for the new atoms. Then, $Q''^{\onDB''}=Q'^{\onDB}$ holds.

Observe now that $Q''$ contains, for each $C_i\in \bar C$, subqueries of the form $\mathit{atom}_{C_i}(\vec{h_{C_i}}) \wedge Q_{C_i}$, where
the latter is the subquery involving all atoms of $Q'$ where some variable from $C_i$ occurs. Because $\mathit{atom}_{C_i}(\vec{h_{C_i}})$ is
sound and complete w.r.t.~to the full query, and all free variables of $Q_{C_i}$ occurs in this atom (by definition of frontier), we have
$\mathit{atom}_{C_i}(\vec{h_{C_i}})^{\onDB''}\subseteq \pi_{\free(Q_{C_i})} (Q_{C_i}^{\onDB''})$. However, this entails that the subquery
$\mathit{atom}_{C_i}(\vec{h_{C_i}}) \wedge Q_{C_i}$ is equivalent to $\mathit{atom}_{C_i}(\vec{h_{C_i}})$, which means that we can actually
drop from $Q''$ all atoms occurring in $Q_{C_i}$, for any $C_i\in\bar C$, without losing anything (we use here the fact that the variables in
$C_i$ do not occur elsewhere in $Q''$). More precisely, the resulting query, say  $Q_f$, does not contain any quantified variable (as all their
components have been deleted) and it is such that $$Q_f^{\onDB''}=\pi_{\free(Q)} (Q'^{\onDB})=\pi_{\free(Q)} (Q_a^{\onDB_a}).$$

Clearly, the tree projection $\HG_a$ for $\HG_{Q'}$ is a tree projection for $\HG_{Q_f}$, too. Indeed, the only additional atoms in $Q_f$ come
precisely from hyperedges of $\HG_a$. Moreover, since $Q_f$ does not contain any quantified variable, the acyclic hypergraph $\HG_a'$ obtained
from $\HG_a$ by removing all (nodes associated with) quantified variables is a tree projection for  $\HG_{Q_f}$.
This concludes the proof, because
$Q_a$ is precisely a query whose associated hypergraph is $\HG_a$, and therefore
from $\HG_a'$ and $\DB_a$ we can obtain a new acyclic equivalent instance $Q_a'$ having no quantified variables (with its associated database
$\DB_a'$). Eventually, recall that the counting problem is well known to be feasible in polynomial time for acyclic queries without quantified variables (see, e.g., ~\cite{PS13}).
\end{proof}

By combining the latter result with the fact that  \sharpdecomposition s can be computed in fixed-parameter polynomial-time (Theorem~\ref{thm:fpt}),
we get that the problem is in FPT, in the general case where arbitrary sets of views are considered.

\begin{cor}\label{thm:fptCounting}
The following problem, parameterized by the query size, is in ${\rm FPT}$:
Given a query $Q$, with a view set $\V$ and a legal database $\DB$,
 solve $\countProblem(Q,\DB)$ if there is a \sharpdecomposition\ of $Q$ w.r.t.~$\V$; otherwise output that
  such  a \sharpdecomposition\ does not exist.
\end{cor}

\section{ {\sf \#}-\emph{hypertree decompositions}}\label{sec:HDs}

As a relevant specialization of the concept of {\sf \#}-covering, we now consider the well-known
\emph{(generalized) hypertree decomposition} method, which is recast in this general framework by considering
the views obtained from all sets of $k$ query atoms, thus obtaining the notion of {\sf \#}-\emph{hypertree decompositions}.

We show that the counting problem is in polynomial-time (not just FPT) for classes of queries having bounded {\sf \#}-\emph{hypertree width},
though both computing generalized hypertree decompositions and computing query cores are $\NP$-hard problems.

 Recall that, for a given query $Q$, $\VQ^{k}$ is the view set built by including, for each
subset $C\subseteq \atoms(Q)$ with $|C| = k$, a fresh atom $w_C$ over the variables $\vars(C)$.
 Moreover, for each relation symbol $r$ occurring in $Q$, $\VQ^{k}$ contains an associated relation symbol $w_{r}$,
 and for each atom $r({\bf X})\in \atoms(Q)$, it contains a view $w_{r}({\bf X})$, called {\em query view}.
  When $Q$ is to be evaluated on a database $\DB$ using $\VQ^{k}$, we need a \emph{legal} database $\DB'$ for the views; this is obtained by initialising query views with the ``input relations'' from $\DB$ and, in the case of hypertree decompositions,
  by initialising every other view in $\VQ^{k}$ with the result of a join operation over a set of $k$ relations from $\DB$.
   Such a database $\DB'$ is called hereafter the {\em standard view extension} of $\DB$ to $\VQ^{k}$.

 It is well-known  and easy to see that every
tree projection of $\HG_Q$ w.r.t.~the hypergraph associated with $\VQ^{k}$ corresponds to a width-$k$ generalized hypertree decomposition of $Q$, and vice versa (see, e.g., \cite{GS17}).
Accordingly, we use the two notions interchangeably.

Recall from Definition~\ref{def:sharpHypertree} that $Q$ has a width-$k$ {\sf \#}-\emph{hypertree decomposition} if it is \sharpcovered\ w.r.t.~$\VQ^{k}$. Its {\sf \#}-\emph{hypertree width} is the minimum width over all its {\sf \#}-hypertree decompositions.
A class $\boA$ has bounded {\sf \#}-\emph{hypertree width} if there exists some finite natural number $k$ such that 
 the {\sf \#}-hypertree width of every query $Q\in\boA$ is at most $k$.

\begin{figure}[h]
 \centering
    \includegraphics[width=0.7\textwidth]{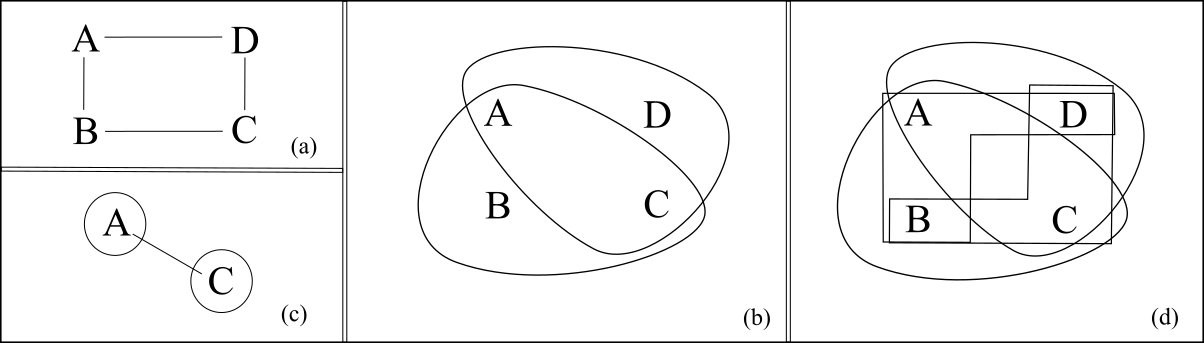}
    \includegraphics[width=0.2\textwidth]{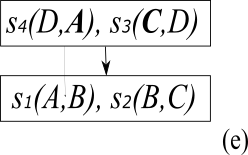}
  \caption{Structures in Example~\ref{ex:esempio3bis}: (a)
  Hypergraph $\HG_{Q_1}$; (b) a \sharpdecomposition\ of $Q_1$ w.r.t.~$\V_{Q_1}^2$; (c) Hypergraph $\FH(Q_1,\free(Q_1))$;  (d) the hypergraph associated with $\V_{Q_1}^2$; (e) a width-2 $\#$-hypertree decomposition of $Q_{1}$.}\label{fig:esempioHypertree}
\end{figure}

\begin{exa}\label{ex:esempio3bis}
Consider the query $Q_1 =\exists B,D\ s_1(A,B)\wedge s_2(B,C)\wedge s_3(C,D)\wedge s_4(D,A)$, whose associated hypergraph $\HG_{Q_1}$ is depicted in Figure~\ref{fig:esempioHypertree}(a). Consider the view set $\V_{Q_1}^k$ with $k=2$ whose associated hypergraph is reported in Figure~\ref{fig:esempioHypertree}(d). These hyperedges represent the views in the standard view extension for $Q_{1}$ built using all pairs of query atoms, that is, the views over the sets of variables
 $\{A,B,C\}$, $\{A,B,D\}$, $\{A,C,D\}$, and $\{B,C,D\}$. 
 Note now that $\free(Q_1)=\{A,C\}$ and that $Q_1$ cannot be simplified, as it is a core. 
 The frontier hypergraph $\FH(Q_1,\free(Q_1))$ is depicted in Figure~\ref{fig:esempioHypertree}(c); observe in particular the hyperedge $\{A,C\}$, which is the frontier of both free variables $A$ and $C$.
 
 The acyclic hypergraph shown in
Figure~\ref{fig:esempioHypertree}(b) is a \sharpdecomposition\ of $Q_1$ w.r.t.~$\V_{Q_1}^2$, because 
 its hyperedges cover both the hyperedges of $\HG_{Q_1}$ and those in the frontier hypergraph.
  Figure~\ref{fig:esempioHypertree}(e) shows a classical representation of
this hypertree decomposition as a join tree, where vertex labels contain the query atoms used to cover the variables occurring in each vertex (equivalently, used to build the views used to deal with those variables).  
 The width of this decomposition is $2$, which is also the {\sf \#}-{hypertree width} of $Q_{1}$, because it is a cyclic core and thus its width is strictly greater than 1.\hfill $\lhd$
\end{exa}

\begin{exa}\label{ex:esempio3bis-bis}
Consider again Example~\ref{ex:esempio1} and recall that Figure~\ref{fig:HD2}(c) shows a width-2
{\sf \#}-{hypertree width} of $Q_0$, which witnesses that its {\sf \#}-{hypertree width} is $2$.
Figure~\ref{fig:esempioCovered}(c) shows its tree projection representation.
\hfill $\lhd$
\end{exa}

In order to get the tractability result,
we first show that a core can be computed in polynomial time if the needed decompositions exist, without the need of computing them.
We use a result on the power of local consistency~\cite{GS17} that
extends a result on hypertree decompositions, based on a different technique~\cite{ChenD05}.

Since we use the general properties of local consistency on tree projections, we remark that this result, which is of independent interest, actually holds for every structural decomposition method whose set of views can be computed in polynomial time, though we state it for hypertree decompositions, for the sake of presentation.

\begin{lem}\label{lemma:core}
Let $k$ be a fixed natural number. Let $Q$ be a query whose cores have (generalized) hypertree width at most $k$.
 Then, any core of $Q$ can be computed in polynomial time.
\end{lem}
\begin{proof}
The proof is based on query minimization techniques already pioneered in \cite{ChandraM1977}.
Observe that if~$Q$ is not a core, then there is a subquery $Q_s$ that we get by deleting an atom from $Q$ that contains a core of $Q$. This (sub)query is homomorphically equivalent to $Q$, in particular, there is a homomorphism from $Q$ to~$Q_s$.  Because all cores are isomorphic, for every substructure $Q_s$ of $Q$ that is homomorphically equivalent to $Q$, the core of $Q_s$ is also a core of $Q$.

Let $\DB_{Q}$ be the database associated with the conjunctive query $Q$ viewed as a relational structure: for each atom $r({\bf X})$ occurring in $Q$, the relation with symbol $r$ in $\DB_{Q}$ contains the tuple of terms {\bf X}.
 Let $\DB^{k}$ be the standard database extension of $\DB_{Q}$ to the view set $\VQ^{k}$, and let $Q_{0}$ be a core of $Q$.

  By hypothesis, there is a tree projection of  $\HG_{Q_0}$ with respect to the hypergraph $\HG^{k}$ associated with $\VQ^{k}$.
  By the results in~\cite{GS17}, for every input database $\DB$, $Q$ has a non-empty answer on $\DB$ if and only if enforcing pairwise consistency on the available views we get a non-empty database. This procedure takes polynomial time in the size of the input database (we compute the join operation on each pair of views, thus removing tuples that do not match, until no more tuple can further be  deleted).

By using this result, the construction of a core $Q_c$ of $Q$ goes as follows. Start with $Q_c$ equals to $Q$. Then,
  take some atom $r({\bf u})$, let $Q_{c}'$ be the query we get from $Q_{c}$ by deleting $r({\bf u})$, let
  $\DB_{Q_{c}'}$ be its associated database, and let $\DB_{Q_{c}'}^{k}$ be its standard extension to the view set $\VQ^{k}$.
  Then, enforce pairwise consistency on $\VQ^{k}$ to decide in polynomial time whether there is some answer on
  $Q$ over $\DB_{Q_{c}'}$, that is, whether there is a homomorphism from $Q$ to $Q_{c}'$.
  If this is the case, the subquery $Q_{c}'$ is homomorphically equivalent to $Q$ and  we set $Q_{c}=Q_{c}'$;
  otherwise, the atom $r({\bf u})$ remains in $Q_{c}$ and we continue by considering another atom for a possible deletion.

 By the above discussion, the minimal subquery $Q_{c}$ obtained at the end of this procedure, which is homomorphically equivalent to the original input query $Q$,  must be one of its cores.
 Note that the overall algorithm takes polynomial-time in the size of $Q$.
 \end{proof}

We next prove the main tractability result.
Recall that, for any given class $\boA$ of queries, $\sCQ[\boA]$ is the problem of computing the number of answers of
 $Q$ on $\DB$, where $\DB$ is any database and $Q$ is any query belonging to $\boA$.
  The input size is the combined size of $Q$ and $\DB$.

\thmtractability*
\begin{proof}
Let $k$ be the maximum {\sf \#}-hypertree width over the queries in $\boA$. We know that it is a finite number, by the hypothesis on the class $\boA$.
We next describe a polynomial-time algorithm $A_{k}$ that computes the correct number of answers for any given instance $(Q,\DB)$,
with $Q\in\boA$.

 Because $Q\in\boA$, this query has a width-$k$ {\sf \#}-hypertree decomposition, thus Algorithm $A_{k}$ first uses
the procedure in Lemma~\ref{lemma:core} to compute a core $Q'$ of $\adorn(Q)$ in polynomial time (in the query size).
 Because  $Q\in\boA$, there exists a width $k$-generalized hypertree decomposition of
 $\HG'=(V,H)$ where $V=\vars(Q')$ and $H= \edges(\HG_{Q'})\cup \edges(\FH(Q',\free(Q)))$.
  Then, from a well known approximation result about hypertree decompositions~\cite{AGG07},
  we can compute in polynomial-time (in the query size) a generalized hypertree decomposition of $\HG'$ having width at most $3k+1$, which has at most $\vars(Q')$ vertices.

  At this point, we have a tree projection (the decomposition hypergraph) that covers both $\adorn(Q)$ and the frontiers hypergraph,
  and we provide a legal database for it by computing the view relation associated with each vertex of the decomposition through a join operation involving at most $3k+1$ atoms. Since $k$ is a constant and the number of vertices is bounded by $\vars(Q')$,
  building this database takes polynomial time.

Finally, having the needed tree projection with the associated relations, Algorithm $A_{k}$ computes the desired number of solutions in polynomial
time, as described in Theorem~\ref{thm:general}.
\end{proof}


\begin{rem}[On Fractional Hypertree Decompositions]
For completeness, note that the same tractability result can be equivalently stated for a wider class of  {\sf \#}-covered queries, based on the notion of fractional hypertree decomposition~\cite{GM06}, rather than generalized hypertree decomposition.
The only difference is that, if $k$ is the maximum width in the considered class of queries,
the view sets used both in Lemma~\ref{lemma:core} and in Theorem~\ref{thm:fptGHTW}
should be defined according to the $O(k^3)$ approximation technique described in~\cite{M09} for computing fractional hypertree decompositions.
\end{rem}

\begin{rem}[On the Quantified Starsize]
 Another approach to counting answers of conjunctive queries with existential variables has been independently proposed by Durand and Mengel in~\cite{DurandM15}.
It is founded on the concept of \emph{quantified star size}, which can be recast, in our notation and terminology, as
the cardinality of the maximum independent set over the frontiers $\Fr(Y,\free(Q),\HG_Q)$, for any quantified variable $Y$, i.e.,
any $Y\in \vars(Q)\setminus\free(Q)$.
It is easy to see that there exist classes of tractable instances
that are not identified using the approach described in~\cite{DurandM15}, but that are indeed identified according to the concept of 
{\sf \#}-hypertree width.
For comparative results between the two concepts, we direct the interested reader to Appendix~\ref{appendix:starsize}. In that section, we also demonstrate how to expand the concept of quantified star size to achieve a similar behaviour as that of {\sf \#}-hypertree decompositions.
\end{rem}

\section{The Frontier of Tractability on Bounded Arity Instances}
\label{sec:frontier}

A problem left open by \cite{DurandM15} is whether the notion of quantified star size (combined with the notion of hypertree width) over classes of
bounded-arity queries precisely characterizes the instances of the counting problem that can be solved in polynomial time.
In particular, it has been shown by \cite{DurandM15} that the property holds over (queries associated with) classes of hypergraphs having bounded
edge-size. However, it turns out that this notion of tractability is not able to identify classes of
queries that are recognized to be tractable according to the notion of \sharpHD\ (see Example~\ref{ex:esempio3} in Appendix~\ref{appendix:starsize}).
In fact, we next show that this latter notion precisely marks the frontier of tractability in the bounded arity case. As for the enumeration
problem, it turns out that dealing with output variables instead of just considering the set of all variables involved in the problem requires
additional technical tools.

The result is a trichotomy in the spirit of the dichotomy stated in the landmark paper by \cite{G07} about the complexity of answering Boolean
conjunctive queries.
However, novel technical tools are needed here to solve the problem.
The following section provides these ingredients.

\input{lowerboundNew}

\section{Hybrid Tractability}\label{sec:ht}

Since we are dealing with a setting where the query and the database are both given as input (neither is constant), it is meaningful to
consider decomposition techniques that are able to exploit structural properties of the query in combination with properties of the given data,
such as functional dependencies or any other feature that may simplify the evaluation. Intuitively, such features may induce different
structural properties that are not identified by the ``worst-possible database'' perspective of purely structural methods. The motivating idea is to end up with algorithms that can be practically applied over a wide range of real world settings where, in particular, such  methods do not suffice alone.

In this section we describe a new hybrid approach where decomposition techniques exploit
information about the database to keep under control critical query fragments via structural techniques.

Recall from the introduction the notion of degree of free variables in a database with respect to a given hypertree:
\begin{defi}\label{def:bound}
Let $\HD=\tuple{T,\chi,\lambda}$ be a hypertree (not necessarily decomposition) of a query $Q$, with free variables $F$, over a database $\DB$.
For a vertex $v$ of $T$, consider the relation $r_{v}$ associated with $v$ according to the hypertree decomposition approach,
that is, $r_v=\ \pi_{\chi(v)}(\bowtie_{q\in \lambda(v)} q^\onDB)$. Consider its projection $r_v[F]=\pi_{F} (r_{v})$ onto the variables in $F$, and let $\deg_{\DB}(F,v) = \max_{\theta\in r_v[F]} ( |\sigma_\theta (r_v)| )$ be the degree of $F$ at vertex $v$.

Then, $\mathit{bound}_{F}(\DB,\HD)$ is the maximum degree of the free variables $F$ across the vertices of the hypertree (typically, we omit the subscript $F$ for simplicity in notation).\hfill $\Box$
\end{defi}

We first show that the algorithm by Pichler and Skritek~\cite{PS13} for acyclic and small-width queries scales exponentially with respect to this degree value only, which in many cases is significantly smaller than the maximum number of tuples over the database relations that is considered in their analysis. 

\begin{restatable}{thm}{thmmainHybrid}\label{thm:mainHybrid}
Let $\HD=\tuple{T,\chi,\lambda}$ be a width-$k$ hypertree decomposition of $Q$, and let $\DB$ be a database such that
$\mathit{bound}(\DB,\HD)\leq h$, and where the maximum number of tuples occurring in its relations is $m$. Then, $\countProblem(Q,\DB)$ can be solved in $O(|\vertices(T)|\times m^{2 k}\times 4^{h})$.
\end{restatable}

    This motivates the question of computing a $\DB$-\emph{optimal width-$k$ hypertree decomposition}, i.e., a decomposition leading to the best possible degree value $\mathit{bound}(\DB,\HD)$.
    We show that this problem is intractable in general, but it is feasible in polynomial time when restricted over classes of hypertree decompositions in \emph{normal form}~\cite{SGL04,GS08}.
    For the sake of readability, these results and the proof of Theorem~\ref{thm:mainHybrid} are reported in Appendix~\ref{sec:bounded}.
    
 Then, we define the new hybrid notion of
  {\sf \#}$_b$-hypertree decomposition, which combines {\sf \#}-coverings and $\DB$-optimal decompositions,
  when low degree values can be attained.
  
\begin{figure}[t]
 \centering
    \includegraphics[width=0.75\textwidth]{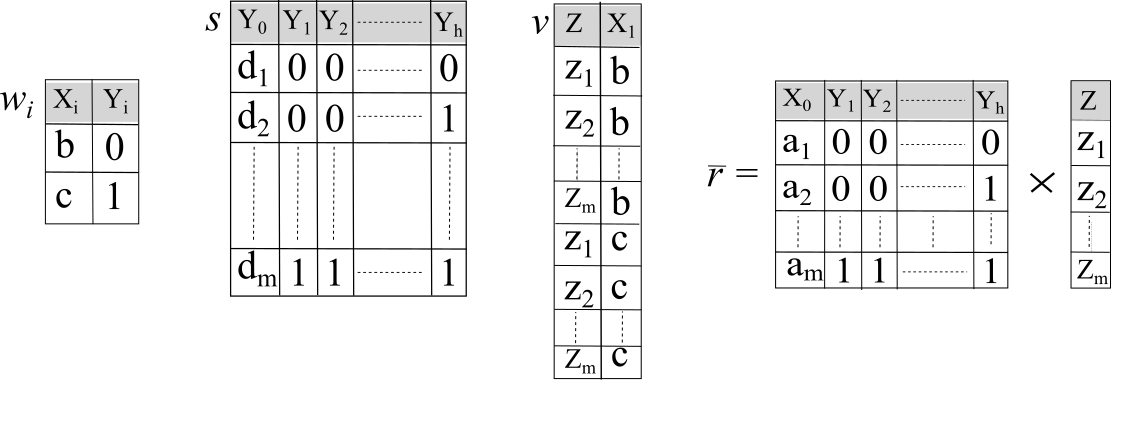}
  \caption{Database in Example~\ref{ex:hybrid3}}\label{fig:hybrid4}
\end{figure}

\begin{figure}[h]
 \centering
    \includegraphics[width=0.47\textwidth]{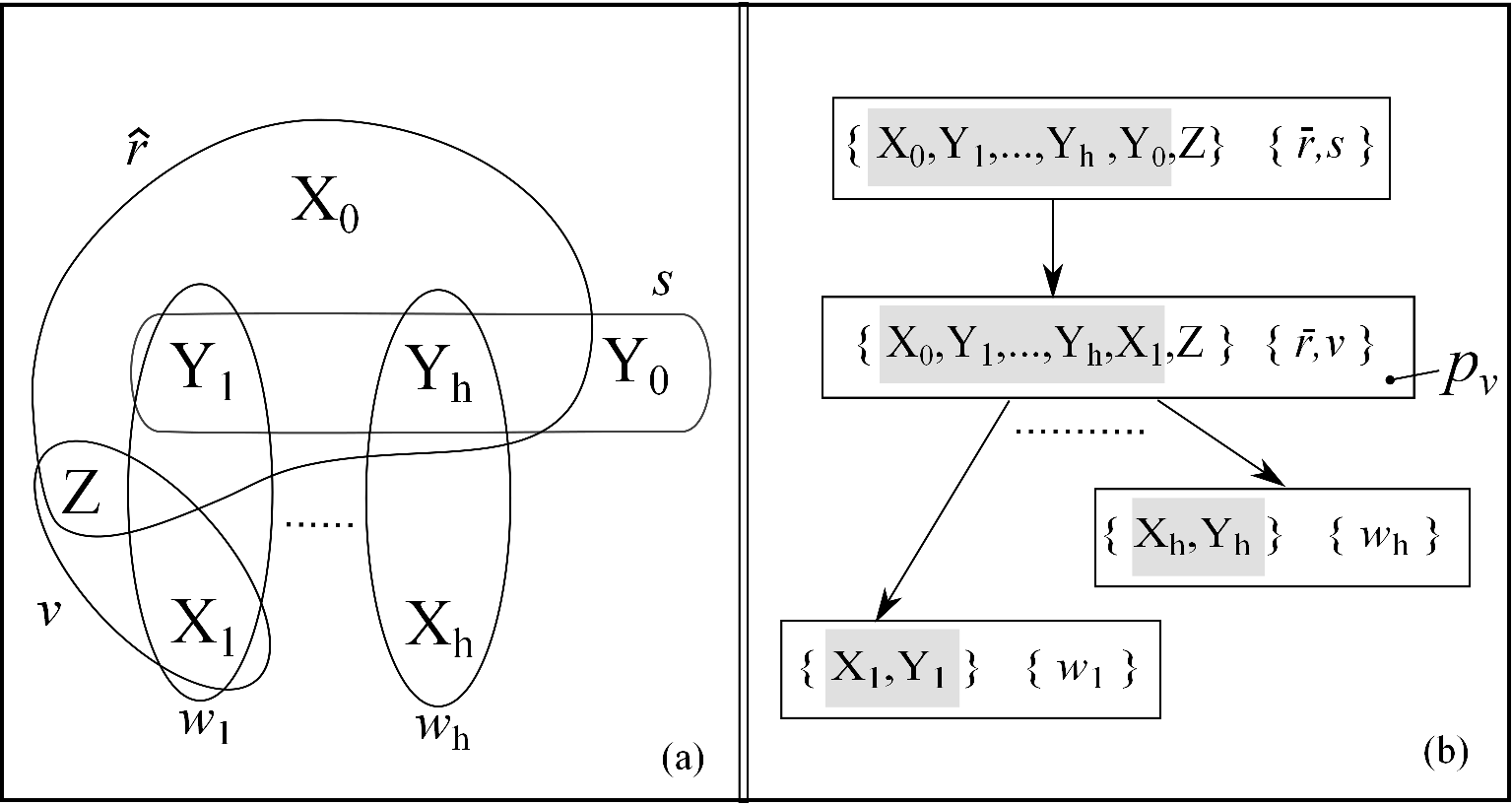}
  \caption{Structures in Example~\ref{ex:hybrid3}: (a) The hypergraph $\HG_{\bar Q_2^h}$; (b) A width-$2$ {\sf \#}$_h$-hypertree decomposition decomposition of $\bar Q^2_h$.}\label{fig:hybrid3}
\end{figure}

\begin{exa}\label{ex:hybrid3}
Consider the class of instances $\{\countProblem(\bar Q_2^{h},\bar \DB_2^m)\}$, for any pair of natural numbers $h$ and $m$, where 
 $\bar \DB_2^m$ is shown in Figure~\ref{fig:hybrid4} and $\bar Q_2^{h}$ is the following query: 
\[\exists Y_0,...,Y_h, Z \quad  \bar r(X_0,Y_1,...,Y_h,Z) \wedge s(Y_0,Y_1,...,Y_h)
                       \bigwedge_{i\in\{1,...,h\}} w_i(X_i,Y_i)\ \wedge v(Z,X_1)\]

By looking at the hypergraph in Figure~\ref{fig:hybrid3}(a), it is easy to see that the frontier of the existential variables is here a clique comprising all the free variables $\{X_{i}\}$, $i\in [h]$.
It follows that this class has no bounded {\sf \#}-generalized hypertree width, and thus the purely structural approach does not work for these instances.

Observe now that in database $\bar \DB_2^m$, in relation $s$ and $\bar r$, the tuples of values of existential variables $\{Y_{i}\}$, $i\in [h]$ encodes the $m = 2^{h}$ numbers from $0$ to $2^{h}-1$. Even the free variables $\{X_{i}\}$, $i\in [h]$ can be assigned in $m$ different ways, however every query answer (assignment on these free variables) has a unique way to be extended to the free variables $\{Y_{i}\}$, $i\in [0,\dots,h]$. On the other hand, every query answer has $m$ possible extensions to variable $Z$, which can be arbitrarily mapped to any value in its domain.
It follows that the degree value $\mathit{bound}(\bar \DB_2^m,\HD)$ is $m$, no matter of the chosen hypertree decomposition $\HD$, and the techniques in Section~\ref{sec:bounded} cannot be applied fruitfully.~\hfill~$\lhd$
\end{exa}

In the example above, unlike variable $Z$, the variables $Y_0, Y_1, ..., Y_h$ exhibit favorable behaviour concerning the potential extensions of their assignments. Hence, they could serve as viable candidates for existential variables to be treated as free ones.

If we assume that $Y_0, Y_1, ..., Y_h$ are treated as free variables, the frontier of $Z$ (the only remaining existential variable) encompasses the set of variables $\{X_{0}, X_{1}, Y_1, ..., Y_h\}$, which coincides with the sole hyperedge of the frontier hypergraph.
We will demonstrate next that we can handle this additional constraint in a structural manner and then count the desired answers by examining a residual query where $Z$ no longer occurs.

In summary, if we identify sets of existential variables that do not pose concerns in terms of their degree values, we can opt to treat them as if they were free variables. By doing so, the connectivity properties of the quantified variables may change, and it's possible that their (new) frontiers may have some guards and a suitable decomposition.
We use a threshold degree-value $b$ that determines to what extent we allow the use of the technique based on the propagation of combinations of assignments of existential variables.

For the purpose of clarity in our presentation, we give the definition of this new concept within the framework and using the notations employed for hypertree decompositions. Nevertheless, extending it to arbitrary view sets and tree projections is straightforward (as for the associated tractability results).

For any set $\bar S\subseteq \vars(Q)$, define $Q[\bar S]$ as the query over the same atoms and variables as $Q$, but where $\free(Q[\bar S])=\bar S$. 

\begin{defi}\label{def:main}
Let $b$ be a fixed natural number, and let  $\HD=\tuple{T,\chi,\lambda}$ be a hypertree of a query $Q$. 
Assume that a set $\bar S\subseteq \vars(Q)$ of variables with
$\free(Q)\subseteq \bar S$ exists such that:

\begin{itemize}
  \item[(1)] $\HD$ is a width-$k$ {\sf \#}-generalized hypertree decomposition of $Q[\bar S]$;

  \item[(2)] $\mathit{bound}_{\free(Q)}(\DB,\tuple{T,\chi_{\bar S},\lambda})\leq b$, where $\chi_{\bar S}(p)=\chi(p)\cap \bar S$.
\end{itemize}

Then, $\tuple{\HD,\bar S}$ is called a width-$k$ {\sf \#}$_b$-generalized hypertree decomposition of $Q$ w.r.t.~$\DB$.\hfill $\Box$
\end{defi}

Condition~(1) requires the existence of a width-$k$ {\sf \#}-generalized hypertree decomposition for a query where all variables in
$\bar S$ are treated as free variables. Thus, variables in $\vars(Q)\setminus \bar S$ are
attacked in a purely structural way, and the frontier of each of them has to be covered in the decomposition. In particular, note that frontiers are
calculated w.r.t.~the core of $\adorn(Q[\bar S])$, that is, the connected components of the quantified variables in the core depend on the
choice of variables in $\bar S$.
Instead, condition (2) is meant to exploit the bounded-degree condition over the query restricted to the variables in $\bar S$. 
In this way, we get a controlled exponential blow-up w.r.t.~$b$ only over these variables (in the example, $Z$ no longer plays any role).


\begin{exa}\label{ex:hybrid4}
Fix the smallest threshold $b=1$. According to the above hybrid notion, the width of all instances $\{\countProblem(\bar Q_2^{h},\bar \DB_2^m)\}$ in Example~\ref{ex:hybrid3}, for any pair of natural numbers $h$ and $m$, is just $2$.

Indeed, for every instance in this class there exists a {\sf \#}$_1$-generalized hypertree decomposition $\tuple{\tuple{T,\chi,\lambda},\bar
S}$ of $\bar Q_2^h$ w.r.t.~$\bar \DB_2^m$ having width $2$. Figure~\ref{fig:hybrid3}(b) shows such a decomposition, where $\bar
S=\free(Q)\cup \{Y_0,Y_1,...,Y_h\}$---the elements in the $\chi_{\bar S}$-labeling are emphasized.
Note that both conditions in Definition~\ref{def:main} are satisfied. Indeed, $Z$ is the only quantified variable that is not in $\bar S$, and
its frontier is covered by a vertex of the decomposition (the vertex $p_v$, in the decomposition tree shown in the figure).

To see that condition (2) holds, note that the degree value is $1$ over $\DB_2$ with respect to this decomposition.
In Figure~\ref{fig:hybrid3}(b), consider the root, say $p_r$, which covers the atoms
defined over $s$ and $\bar r$. Because $Z\notin \chi_{\bar S}(p_r)=\chi(p_r)\cap \bar S$, it will be projected out when this vertex will be
evaluated. Then, without $Z$, $X_0$ acts as a key for the relation computed for evaluating $p_r$. 
The same occurs at vertex $p_v$, where again $Z$ does not occur in the $\chi_{\bar S}$-labeling.
Likewise, at the remaining vertices covering atoms defined over $w_{i}$, the free variables $X_{i}$ act as keys.
\hfill $\lhd$
\end{exa}

By suitably combining the results in the previous section, we can show that the above hybrid notion is able to guarantee the tractability of the counting problem.

\begin{thm}\label{thm:generalH}
Let $k$ and $b$ be fixed natural numbers. If a width-$k$ {\sf \#}$_b$-generalized hypertree decomposition $\HD$ of $Q$ w.r.t.~$\DB$ is given, then
$\countProblem(Q,\DB)$ can be solved in polynomial time (w.r.t. $||Q||$, $||\DB||$, and $||\HD||$).
\end{thm}
\begin{proof}
Let $\tuple{\HD,\bar S}$, with $\HD=\tuple{T,\chi,\lambda}$, be a width-$k$ {\sf \#}$_b$-hypertree decomposition of $Q$ w.r.t.~$\DB$. Recall
from condition (1) in Definition~\ref{def:main} that $\HD$ (more formally, its hypergraph $\HG_a$ with $\edges(\HG_a)=\{ \chi(p) \mid p\in
\vertices(T) \}$) is a width-$k$ {\sf \#}-generalized hypertree decomposition of $Q[\bar S]$.

Recall that, by definition, $\free(Q[\bar S])=\bar S$. We apply the algorithm in the proof of Theorem~\ref{thm:general} to $Q[\bar S]$ and
$\HG_a$. So, in polynomial time, we can end up with a solution equivalent query ${Q_f}$ without any quantified variable in $Q[\bar S]$ and with
its associated database $\DB_f$. In particular, $Q_f$ is obtained from $Q$ by deleting all atoms involving variables in $\vars(Q)\setminus \bar
S$, and by adding fresh atoms of the form $atom(\bar X_Y)$, with $\bar X_Y= \Fr(Y, \bar S, \HG_{Q'[\bar S]})$ for each quantified variable
$Y\in\vars(Q')\setminus \bar S$, where $Q'$ is some core of $\adorn(Q)$.

Consider now a hypertree $\bar \HD=\tuple{T,\chi_{\bar S},\lambda}$ that is the same as $\HD$ but for the labeling  $\chi_{\bar
S}(p)=\chi(p)\cap \bar S$. Note that $\bar \HD$ satisfies conditions (1), (2), and (3) in the definition of a width-$k$ hypertree decomposition
of $Q_f$. Condition (4) might be violated because it is a generalized hypertree decomposition, but it is immediate to check that this condition
does not play any role in the algorithm in Figure~\ref{fig:algoritmo}, which only relies on the fact that the labeling $\chi_{\bar S}$
identifies an acyclic hypergraph, so that we can apply the arguments by Pichler and Skritek~\cite{PS13}. Therefore, we use
Theorem~\ref{thm:mainHybrid} on $Q_f$, $\DB_f$, and the hypertree $\bar\HD$.
Observe that variables in $Q_f$ are quantified as in $Q$, so that in particular each $Y\in \bar S\setminus\free(Q)$ is treated according to its
original role of existential variable. Thus, we derive that the counting problem can be solved in polynomial time, whenever the degree value
$\mathit{bound}(\DB_f,\tuple{T,\chi_{\bar S},\lambda})$ is less than some fixed constant. This is guaranteed by Condition~(2) in
Definition~\ref{def:main} and by the fact that $\DB_f$ is computed by enforcing pairwise consistency over views built using at most $k$
relations from $\DB$. As a consequence, for every $v$ in $T$, $\pi_{\chi_{\bar S}(v)}(\bowtie_{q\in \lambda(v)} q^{\onDB_f})\subseteq
\pi_{\chi_{\bar S}(v)}(\bowtie_{q\in \lambda(v)} q^\onDB)$.
\end{proof}

This result can be useful in many practical applications. As suggested by Example~\ref{ex:hybrid4}, a noticeable case is when some functional
dependencies occur in the database. For instance, if some (often all) existential variables are functionally determined by keys (over which we want to count the solutions), then their boundedness value is always $1$, so that the technique may freely use them as if they were free variables, if this is convenient to decompose the given query. 
The approach is also effective in the presence of \emph{quasi-keys}, i.e., attributes whose values identify a small subset of all the substitutions in a relation, or more generally in the presence of degree-constraints.

Our second result is that computing an {\em optimal} hybrid decomposition is fixed-parameter tractable.

\begin{thm}\label{thm:fptH}
Let $k$ be a fixed natural number. The following problem, parameterized by the query size, is in ${\rm FPT}$: Given a query $Q$ and a database
$\DB$, compute a width-$k$ {\sf \#}$_b$-generalized hypertree decomposition of $Q$ w.r.t.~$\DB$ (if any), 
having the minimum degree value.
\end{thm}
\begin{proof}
Let $k$ be a fixed natural number. Given a query $Q$ on a database $\DB$, we first build the view set $\VQ^k$ associated with the generalized
hypertree decomposition method, with its database $\DB^k$. This is feasible in polynomial time. Let $m$ be the size of the largest relation in
$\DB^k$.

Then, for every $h:= 1,\dots, m$ and every subset $\bar S$ that includes $\free(Q)$ we perform the following procedure. First, we compute a
core $Q'$ of $\adorn(Q[\bar S])$. Consider the hypergraph $\HG'=(V,H)$ where $V=\vars(Q[\bar S])$ and $H= \edges(\HG_{Q'})\cup
\edges(\FH(Q',\free(Q[\bar S])))$. If $\HG'\leq \HG_{\VQ^k}$, then we have a chance to find a decomposition, but we have to keep the
boundedness value under control. Therefore, we build the following view set  $\VQ^{k,h}[\bar S]$ with its associated database $\DB^{k,h}[\bar
S]$: for each view $w'\in \VQ^k$ such that $\vars(w')\cap \bar S\neq \emptyset$, the view set contains a view $w$ with $\vars(w)= \vars(w')\cap
\bar S$ and the database contains an associated relation $r_w= \pi_{\vars(w)} (w'^{\DB^k})$, if $\forall \theta\in \pi_{\free(Q)}(r_w)$,
$|\sigma_\theta (r_w)|\leq h$. Then, we try to compute in {\rm FPT} a tree projection of $\HG'$ w.r.t. the hypergraph associated with
$\VQ^{k,h}[\bar S]$. If we succeed, we have computed a width-$k$ {\sf \#}$_h$-generalized hypertree decomposition of $Q$ where $h$ is the
minimum possible value, and we halt the procedure. Otherwise (in case of failure), we try with another set $\bar S$ or with the next value for
$h$, if no further choices are available for $\bar S$. Note that all this loop is feasible in time $f(||Q||)\times poly(||Q||+||\DB||)$, where
 $f(\cdot)$ is a function exponential in the fixed parameter $||Q||$.
\end{proof}

Let $\boADB$ be a class of pairs $(Q,\DB)$, where $Q$ is a query and $\DB$ a database. We say that $\boADB$ has bounded \emph{hybrid
generalized hypertree width} if there are two finite numbers $k$ and $b$ such that, for each $(Q,\DB)\in \boADB$, there is a width-$k$ {\sf
\#$_b$}-generalized hypertree decomposition of $Q$ w.r.t.~$\DB$.
For instance, this is the case of the class in Example~\ref{ex:hybrid4}, for $b=1$ and $k=2$. 
From the above results, we immediately get that the counting problem
is fixed parameter tractable for such classes.

\begin{cor}
Let $\boADB$ be a class having bounded hybrid generalized hypertree width. Then, for each $(Q,\DB)\in\boADB$, the problem
$\countProblem(Q,\DB)$, parameterized by the query size, is in ${\rm FPT}$.
\end{cor}

\section{Conclusion}\label{sec:conclusion}

In the paper, we have studied islands of tractability for the problem of counting answers to conjunctive queries, from the classical
purely-structural perspective as well as from a hybrid perspective.

We explored the combined complexity of the problem, as well as its parameterized complexity, where the size of the query serves as the fixed parameter.
We made no assumptions about the bound on the arities of relations, and we considered the extensive framework of tree projections, encompassing most structural decomposition methods. While our focus was on (generalized) hypertree decompositions, it's noteworthy that all results apply to fractional hypertree decompositions as well.

In the context of instances with bounded arities, we solved the open question concerning the tractability frontier. Our findings reveal that the tractable classes are precisely those exhibiting bounded {\sf \#}-hypertree width.

The tractability frontier for the broader scenario involving classes with unbounded arities  is yet to be explored.
A potential strategy to tackle this challenge could involve utilizing the \#-{\rm subm} counting extension of the submodular width~\cite{FAQ-subm}, along with our methodology of considering colored cores. Additionally, leveraging our case complexity framework could be instrumental in establishing the lower bound.

Concerning the hybrid decomposition methods, a natural avenue of further research is to
incorporate them within concrete query optimizers (as in \cite{GGGS07}).
In principle, our hybrid notions could also be adapted for seamless integration into the algorithms described in~\cite{FAQ-ngo}, allowing us to consider a wider range of "feasible" orderings.

\bibliographystyle{alphaurl}

\bibliography{counting}

\input{appendixNew}

\end{document}

%% file: lowerboundNew.tex

\subsection{Case Complexity}
\label{sct:casecomplexity}

In this section we develop a version of the case complexity framework
advocated in \cite{Chen14} which is 
suitable for classifying counting problems.
A main motivation for this framework is the growing amount of research on parameterized problems which are restricted by the permitted values of the parameter. In particular, this kind of problem arises naturally in query answering problems where one often restricts the admissible queries for the inputs (see e.g.~\cite{G07,DJ04,Chen14}). An aim of the case complexity framework as introduced in \cite{Chen14} is to facilitate reductions between  the considered restricted parameterized problems and to show results independent of computability assumptions for the parameter. 

Throughout, we use $\powfin(S)$ to denote the set containing each finite subset of $S$, that is, the restriction of the power set of $S$ to finite sets.

The central notion for our framework is the following: A \emph{case problem} consists of a problem
$Q: \Sigma^* \times \Sigma^* \to \N$
and a subset $S \subseteq \Sigma^*$,
and is denoted $Q[S]$.
When $Q[S]$ is a case problem, we define the following:
\begin{itemize}

\item $\param{Q[S]}$ is the parameterized problem $(P, \pi_1)$
where $P(s, x)$ is defined as equal to 
$Q(s, x)$ if $s \in S$, and as $0$ otherwise.

\item $\prom{Q[S]}$ is the promise problem $\lpr Q, S \times \Sigma^* \rpr$.

\item $\paramprom{Q[S]}$ is the parameterized promise problem
$(\prom{Q[S]}, \pi_1)$.

\end{itemize}

The case problem we consider in this paper will nearly exclusively be $\sCQ[\C]$ where $\C$ is a class a class of conjunctive queries. Nevertheless, we stress the fact that our framework is fully generic and we believe that it will in the future also be useful for presenting and proving complexity classifications for other problems.

We now introduce a reduction notion for case problems.

\begin{defi}
A \emph{counting slice reduction} from a case problem $Q[S]$ to a second case problem $Q'[S']$ consists of
\begin{itemize}
 \item a computably enumerable language $U\subseteq \Sigma^* \times \powfin(\Sigma^*)$, and
 \item a partial function 
$r: \Sigma^* \times \powfin(\Sigma^*) \times \Sigma^*\rightarrow \Sigma^*$ 
that has domain $U\times \Sigma^*$ and is 
FPT-computable with respect to $(\pi_1, \pi_2)$ via
an algorithm $A$ that, on input $(s, T, y)$, 
may make queries of the form $Q'(t,z)$
where $t \in T$,
\end{itemize}
such that the following conditions hold:
\begin{itemize}
 \item (coverage) for each $s\in S$, there exists $T \subseteq S'$
such that $(s, T)\in U$, and 
 \item (correctness) for each $(s, T) \in U$, it holds (for each $y\in \Sigma^*$) that 
 \[Q(s,y) = r(s, T, y).\]
\end{itemize}
\end{defi}

As usual in counting complexity, it will often not be necessary to use the full generality of counting slice reductions. Therefore, we introduce a second, parsimonious notion of reductions for case problems which is often general enough but easier to deal with.

\begin{defi}
A \emph{parsimonious slice reduction} from a case problem $Q[S]$ to a second case problem $Q'[S']$ consists of
\begin{itemize}
 \item a computably enumerable language $U\subseteq \Sigma^* \times \Sigma^*$, and
 \item a partial function 
$r: \Sigma^* \times \Sigma^* \times \Sigma^*\rightarrow \Sigma^*$ 
that has domain $U\times \Sigma^*$ and is 
FPT-computable with respect to $(\pi_1, \pi_2)$ 
\end{itemize}
such that the following conditions hold:
\begin{itemize}
 \item (coverage) for each $s\in S$, there exists $s'\in S'$ such that $(s, s')\in U$, and 
 \item (correctness) for each $(t, t') \in U$, it holds (for each $y\in \Sigma^*$) that 
 \[Q(t,y) = Q'(t', r(t, t', y)).\]
\end{itemize}
\end{defi}

We give some basic properties of counting slice reductions. 
Their proofs can be found in Appendix~\ref{app:case}.

\begin{prop}
If $Q[S]$ parsimoniously slice reduces to $Q'[S']$, then 
 $Q[S]$ counting slice reduces to $Q'[S']$.
\end{prop}

\begin{thm} \label{thm:transitivity}
Counting slice reducibility is transitive.
\end{thm}

The next two theorems give the connection between case complexity and parameterized complexity. 
In particular, they
show that, from a counting slice reduction,
one can obtain 
complexity results for the corresponding
parameterized problems.

\begin{thm}
\label{thm:slice-red-gives-fpt-red}
Let $Q[S]$ and $Q'[S']$ be case problems.
Suppose that $Q[S]$ counting slice reduces to $Q'[S']$,
and that both $S$ and $S'$ are computable.
Then $\param{Q[S]}$ counting FPT-reduces to $\param{Q'[S']}$.
\end{thm}


\begin{thm}
\label{thm:promfpt-to-fpt}
Let $Q[S]$ be a case problem, and let
$K: \Sigma^* \times \Sigma^* \to \N$ be a problem.
Suppose that $\paramprom{Q[S]}$ is in $\prom{\FPT}$,
$S$ is computably enumerable, and that
the case problem $K[\Sigma^*]$ counting slice reduces to $Q[S]$.
Then the parameterized problem $(K, \pi_1)$ is in $\FPT$.
\end{thm}

In the remainder of the paper, we will show all our reductions in the case complexity framework and then use Theorem~\ref{thm:slice-red-gives-fpt-red} and Theorem~\ref{thm:promfpt-to-fpt} to derive parameterized complexity results. This approach lets us give results on $\sCQ[\C]$ for different complexity assumptions on $\C$ without having to deal with these assumptions in the proofs. Thus we separate the technicalities of the reductions from the assumptions on $\C$ which in our view gives a far clearer presentation.

\subsection{The Trichotomy}

Let $\boA$ be a class of queries. Recall that we denote by $\Frontiers(\boA)$ the frontier hypergraphs of these queries, that is,
  $\Frontiers(\boA) = \{ \FH(Q',\free(Q)) \mid  Q\in\boA $ and $Q'$ is some core of $\adorn(Q)\}$.
  
In this section we prove the main classification result of this paper, stated in the introduction:
\thmtrichotomy*

The tractability part of Theorem~\ref{thm:real-trichotomy} follows from Theorem~\ref{thm:fptGHTW}.
The other two parts of
Theorem~\ref{thm:real-trichotomy} follow from the following theorem, phrased in terms of case complexity, along with
Theorem~\ref{thm:slice-red-gives-fpt-red}.

\begin{thm} \label{thm:trichotomy}
Let $\boA$ be a computable class of bounded-arity queries.
\begin{enumerate}
 \item If the queries in $\boA$ have unbounded {\sf \#}-hypertree width, but the hypergraphs
  $\Frontiers(\boA)$  have bounded hypertree width, then $\sCQ[\boA]$ is equivalent to $\clique[\mathbb{N}]$ under counting slice reductions.

 \item Otherwise, there is a counting slice reduction from $\sclique[\mathbb{N}]$ to $\sCQ[\boA]$.
\end{enumerate}
\end{thm}

The following lemma directly proves Theorem~\ref{thm:trichotomy}.

\begin{lem}\label{lem:hardness}
Let $\boA$ be a computable class of bounded-arity queries having unbounded {\sf \#}-hypertree width.
\begin{enumerate}
\item If the hypergraphs $\Frontiers(\boA)$  have bounded hypertree width, then there is a counting slice reduction from $\sclique[\mathbb{N}]$ to $\sCQ[\boA]$.
\item  otherwise, $\sCQ[\boA]$ is equivalent to $\clique[\mathbb{N}]$, with respect to counting slice reductions.
\end{enumerate}
\end{lem}

In the remainder of this section, we will show Lemma~\ref{lem:hardness} in several steps which we present in individual subsections.

\subsection{Simulating unary relations}

In this section we show that for queries whose coloring is a core we can simulate unary atoms on the variables of the query. These additional atoms will later allow us to tell the variables apart such that we can later simulate the case in which all atoms of the queries have different relation symbols.

 \begin{lem}\label{lem:bijection}
  Let $Q$ be a conjunctive query such that $\adorn(Q)$ is a core. Then every homomorphism $h:Q\rightarrow Q$ with $h|_{\free(Q)} = \id$ is a bijection.
 \end{lem}
 
\begin{proof}
 Clearly, $h$ is also a homomorphism $h:\adorn(Q)\rightarrow \adorn(Q)$, because $h(a)=a\in r _a^{\adorn(Q)}$ for every $a\in \free(Q)$. But by assumption $\adorn(Q)$ is a core, so there is no homomorphism from $\adorn(Q)$ to a proper substructure and thus $h$ must be a bijection on $\adorn(Q)$ and consequently also on $Q$.
\end{proof}

We assign to every query $Q$ a query $\fadorn(Q)$ which we call the \emph{full coloring of $Q$}: To get $\fadorn(Q)$ we add to $Q$ an atom $r_X(X)$ for \emph{every} variable $X$ of $Q$. 

Note that $\adorn(Q)$ and $\fadorn(Q)$ differ in which atoms we add: For the structure $\adorn(Q)$ we add~$r_X(X)$ only for the free variables of $Q$ while for $\fadorn(Q)$ we add $r_X(X)$ for \emph{all} variables. Thus,~$\fadorn(Q)$ in general may have more atoms than $\adorn(Q)$.

We now formulate the main lemma of this section whose proof uses ideas from \cite{DJ04}.

\begin{lem}\label{lem:constants}
Let $\boA$ be a class of conjunctive queries such that for each $Q\in \boA$ the coloring $\adorn(Q)$ is a core. Let $\boA^*:=\{\fadorn(Q) \mid Q\in \boA\}$. Then there is a counting slice reduction from $p$-$\sCQ[\boA^*]$ to $p$-$\sCQ[\boA]$.
\end{lem}
\begin{proof}
For every query $Q$ from $\boA$, the relation $U$ of our counting slice reduction contains $(\fadorn(Q), Q)$.
Obviously, $U$ is computable and satisfies the coverage property.

 Let $(\fadorn(Q),\relB)$ be an input for $\sCQ[\boA^*]$. Remember that $\fadorn(Q)$ and $\relB$ are over the vocabulary $\tau_Q \cup \{r_X\mid X\in \vars(Q)\}$. 
 We will reduce the computation of the size $|\fadorn(Q)(\relB)|$ to the computation of $|Q(\relB')|$ for different structures $\relB'$.
 
 Let $D:= \{(X,b)\in \vars(Q)\times B\mid b \in R_X^\relB\}$ and define a structure $\DB$ over the vocabulary $\tau_Q$ with the domain $D$ that contains for each relation symbol $r\in \tau$ the relation 
 \begin{align*}r^\DB:= \{ ((X_1, b_1), \ldots , (X_\ell, b_\ell)) \mid &(X_1, \ldots , X_\ell)\in r^Q, (b_1, \ldots , b_\ell)\in r^\relB, \\&\forall i\in [\ell]: (X_i, b_i)\in D\}.\end{align*} 
 Note that here and in the remainder of the proof we are relying on the perspective of seeing $Q$ as a relational structure itself as described in the preliminaries.
 
 Let $\pi_1: D\rightarrow \vars(Q)$ be the projection onto the first coordinate, i.e., $\pi_1(X,b):=X$.
 Observe that $\pi_1$ is by construction of $\DB$ a homomorphism from $\DB$ to $Q$. For every mapping $h: \vars(Q) \rightarrow D$, we define the mapping $(\pi_1 \circ h): \vars(Q) \rightarrow \vars(Q)$ by setting for every $X\in \vars(Q)$ as $(\pi_1\circ h)(X) := \pi_1(h(X))$.
 
 We will use the following claim several times:
 
 \begin{claim}\label{clm:automorphism}
  Let $h$ be a homomorphism from $Q$ to $\DB$ with $(\pi_1\circ h)(\free(Q))=\free(Q)$. Then $\pi_1\circ h$ is an automorphism of $Q$.
 \end{claim}
\begin{proof}
 Let $g:=\pi_1\circ h$. As the composition of two homomorphisms, $g$ is a homomorphism from $Q$ to $Q$. Furthermore, by assumption $g|_{\free(Q)}$ is a bijection from $\free(Q)$ to $\free(Q)$. Since $\free(Q)$ is finite, there is $i\in \mathbb{N}$ such that $g^i|_{\free(Q)}= \id$. But $g^i$ is a homomorphism and thus, by Lemma \ref{lem:bijection}, $g^i$ is a bijection. It follows that $g$ is a bijection. 
 
Since $\vars(Q)$ is finite, there is $j\in \mathbb{N}$ such that $g^{-1}= g^j$. It follows that~$g^{-1}$ is a homomorphism and thus $g$ is an automporphism.
\end{proof}
 
 Let $\calN$ be the set of mappings $h:\free(Q)\rightarrow D$ with $\pi_1 \circ h = \id$ that can be extended to a homomorphism $h':Q\rightarrow \DB$.
 
 \begin{claim}\label{clm:constIdentity}
There is a bijection between $\fadorn(Q)(\relB)$ and $\calN$.
 \end{claim}
\begin{proof}
For each $h^*\in \fadorn(Q)(\relB)$ we define $P(h^*):=h$ by $h(X):=(X,h^*(X))$ for $X\in \free(Q)$. From the extension of $h^*$ to $\vars(Q)$ we get an extension of $h$ that is a homomorphism and thus $h\in \calN$. Thus $P$ is a mapping $P:\fadorn(Q)(\relB)\rightarrow \calN$.

We claim that $P$ is a bijection. Clearly, $P$ is injective. We we will show that it is surjective as well. To this end, let $h:\free(Q)\rightarrow D$ be a mapping in $\calN$ and let~$h_e$ be a homomorphism from $Q$ to $\DB$ that is an extension of $h$. By definition of~$\calN$ such a $h_e$ must exist. By Claim \ref{clm:automorphism} we have that $\pi_1\circ h_e$ is an automorphism, and thus $(\pi_1\circ h_e)^{-1}$ is a homomorphism. We set $h_e' := h_e\circ(\pi_1\circ h_e)^{-1}$. Obviously,~$h_e'$ is a homomorphism from $Q$ to $\DB$, because $h_e'$ is the composition of two homomorphisms. Furthermore, for all $X\in \free(Q)$ we have $h_e'(X)= (h_e\circ (\pi_1\circ h_e))(X) = (h_e\circ (\pi_1\circ h))(X) = h_e(X) = h(X)$, so $h_e'$ is an extension of $h$. Moreover $\pi_1\circ h_e' = (\pi_1\circ h_e)\circ (\pi_1\circ h_e)^{-1} = \id$. Hence, we have $h_e' = \id \times \hat{h}$ for a homomorphism $\hat{h}:Q \rightarrow {\hat{\relB}}$, where $\hat{\relB}$ is the structure we get from $\relB$ by deleting the relations~$r_X^{\relB}$ for $X\in \vars(Q)$. But by definition $h_e'(X)\in D$ for all $X\in \vars(Q)$ and thus $\hat{h}(X)\in R_X^\relB$. It follows that $\hat{h}$ is a homomorphism from $Q$ to $\relB$. We set $h^*:=\hat{h}|_{\free(Q)}$. Clearly, $h^*\in Q(\relB)$ and $P(h^*) = h$. It follows that $P$ is surjective. This proves the claim.
\end{proof}

Let $I$ be the set of mappings $g:\free(Q)\rightarrow \free(Q)$ that can be extended to an automorphism of $Q$. Let $\calN'$ be the set of mappings $h:\free(Q) \rightarrow D$ with $(\pi_1 \circ h)(\free(Q))=\free(Q)$ that can be extended to homomorphisms $h':Q\rightarrow \DB$.

\begin{claim}
 \[|\fadorn(Q)(\relB)| = \frac{|\calN'|}{|I|}.\]
\end{claim}
\begin{proof}
Because of Claim \ref{clm:constIdentity} it is sufficient to show that
\begin{equation}\label{eq:0a}
|\calN'| = |\calN||I|. 
\end{equation}
We first prove that
\begin{equation}\label{eq:1a}
\calN' = \{f\circ g \mid f\in \calN, g\in I\}. 
\end{equation}
The $\supseteq$ direction is obvious. For the other direction let $h\in \calN'$. Let~$h'$ be the extension of $h$ that is~a homomorphism $h':Q\rightarrow \DB$. By Claim \ref{clm:automorphism}, we have that $g := \pi_1\circ h'$ is an automorphism of $Q$. It follows that $g^{-1}|_{\free(Q)}\in I$. Furthermore, $h\circ g^{-1}|_{\free(Q)}$ is a mapping from $\free(Q)$ to~$D$ and $h'\circ g^{-1}$ is an extension that is a homomorphism from $Q$ to $\DB$. Furthermore $(\pi_1\circ h'\circ g^{-1}|_{\free(Q)})(X) = (g|_{\free(Q)} \circ g^{-1}|_{\free(Q)})(X)=X$ for every $X\in {\free(Q)}$ and hence $h'\circ g^{-1}|_{\free(Q)} \in \calN$ and $h= h \circ g^{-1}|_{\free(Q)} \circ g|_{\free(Q)}$ which proves the claim~(\ref{eq:1a}).

To show (\ref{eq:0a}), we claim that for every $f,f'\in \calN$
and every $g, g'\in I$, if $f\ne f'$ or $g\ne g'$, then $f\circ g\ne f'\circ g'$. To see this, observe that $f$ can always be written as $f= \id\times f_2$ and thus $(f\circ g)(X) = (g(X), f_2(g(X))$. Thus, if $g$ and $g'$ differ, $\pi_1\circ f\circ g \ne \pi_1 \circ f' \circ g'$ and thus $f\circ g\ne f'\circ g'$. Also, if $g=g'$ and $f\ne f'$, then clearly $f\circ g\ne f'\circ g'$. This completes the proof of (\ref{eq:0a}) and the claim.
\end{proof}

Clearly, the set $I$ depends only on $Q$ and thus it can be computed by an $\FPT$-algorithm. Thus it suffices to show how to compute $|\calN'|$ in the remainder of the proof.

For each set $T\subseteq \free(Q)$ we define $\calN_T:= \{h\in Q(\DB) \mid (\pi_1\circ h)(\free(Q)) \subseteq T\}$. We have by inclusion-exclusion 
\begin{equation}\label{eq:2a}
|\calN'| = \sum_{T\subseteq \free(Q)} (-1)^{|\free(Q)\setminus T|} |\calN_T|.                                                                                                                        
\end{equation}
Observe that there are only $2^{|\free(Q)|}$ summands in (\ref{eq:2a}) and thus if we can reduce all of them to $\sCQ$ with the query $Q$ this will give us the desired counting slice reduction.

We will now show how to compute the $|\calN_T|$ by interpolation.
So fix a $T\subseteq \free(Q)$. Let $\calN_{T,i}$ for $i=0,\ldots, |\free(Q)|$ 
consist of the mappings $h\in Q(\DB)$ such that there are exactly~$i$ elements $X\in \free(Q)$ that are mapped to $h(X)= (X', b)$ such that $X'\in T$.
Obviously, $\calN_T = \calN_{T,|\free(Q)|}$ with this notation.

Now for each $j=1,\ldots , |\free(Q)|$ we construct a new structure $\DB_{j,T}$ over the domain $D_{j,T}$. 
To this end, for each $X \in T$, let $X^{(1)}, \ldots, X^{(j)}$ be copies of $X$
which are not in $D$. Then we set \begin{align*}D_{j,T}:= \{(X^{(k)}, b)\mid (X,b)\in D, a\in T, k\in [j]\} \cup \{(X,b)\mid (X,b)\in D, X\notin T\}.\end{align*} We define a mapping $B:D\rightarrow \wp(D_{j,T})$, where $\wp(D_{j,T})$ is the power set of $D_{j,T}$, by
\[B(a,b):= \begin{cases}
            \{(X^{(k)},b)\mid k\in [j]\}\}, & \text{if } a\in T\\
            \{(X,b)\}, & \text{otherwise}.
           \end{cases}\]
For every relation symbol $r\in \tau$ we define
$r^{\DB_{T,j}} := \bigcup_{(d_1, \ldots, d_s)\in R^{D}} B(d_1)\times \ldots \times B(d_s).$

Then every $h\in \calN_{T,i}$ corresponds to $i^j$ mappings in $Q(\DB_{j,T})$. Thus for each~$j$ we get 
$\sum_{i=1}^{|\free(Q)|} i^j |\calN_{T,i}| = |Q\DB_{j,T})|.$
This is a linear system of equations and the corresponding matrix is a Vandermonde matrix, so $\calN_T = \calN_{T,|\free(Q)|}$ can be computed with an oracle for $\sCQ$ on the instances $(Q,\DB_{j, T})$. The size of the linear system depends only on~$|\free(Q)|$. Furthermore, $\|D^j\|\le \|D\| j^s\le \|D\|^{s+1}$ where $s$ is the bound on the arity of the relations symbols in $\tau_Q$ and thus a constant. It follows that the algorithm described above is a counting slice reduction. This completes the proof of Lemma~\ref{lem:constants}. $\Box$
\end{proof}

\subsection{Reducing from simple queries to general queries}

In this section we show that we can in certain situations reduce from $\sCQ$ on simple queries to $\sCQ$ on general queries. This will later allow us to reduce from known results from~\cite{DurandM15} to show the hardness results of Lemma~\ref{lem:hardness}.

To every query $Q$ we associate a simple query $Q'$ by iteratively renaming some of its relation symbols until the query becomes simple. Note that up to the names of the relation symbols, the query $Q'$ is unique, so we talk of \emph{the} simple query of $Q$ and denote it by $\simple(Q)$. For a class of $\boA$ of queries we define $\simple(\boA)$ to be the class containing the simple queries of all queries in $\boA$.

We will use the following classical result.

\begin{thm}[\cite{ChandraM1977}]\label{thm:ChandraMerlin}
Let $Q_1, Q_2$ be two conjunctive queries such that their colorings have the same core, then $Q_1$ and $Q_2$ are logically equivalent.
\end{thm}

We proceed in several steps.
Let in this section $\boA$ be a class of conjunctive queries of bounded arity. 
To every query $Q$ we construct a query $\hat{Q}$ as follows: Let $Q'$ be the core of the coloring of $Q$. 
 Then we define $\hat{Q}$ to be the query that we get by deleting the relations~$r_X$ for $X\in \free(Q)$ that we added in the construction of $\adorn(Q)$. We set $\hat{\boA}:=\{\hat{Q} \mid Q\in \boA\}.$

\begin{claim}\label{clm:hattowithouthat}
  There is a parsimonious slice-reduction from $\sCQ[\hat{\boA}]$ to $\sCQ[\boA]$.
 \end{claim}
\begin{proof}
The relation $U$ relates to every query $Q$ the query $\hat{Q}$. Certainly, $U$ is computable and by definition assigns to each query in $\hat{\boA}$ a query in $\boA$.
 
Note that, by construction, the cores of $\adorn(Q)$ and $\adorn(\hat{Q})$ are identical.
 Hence, by Theorem \ref{thm:ChandraMerlin} we have that $Q$ and $\hat{Q}$ are 
 logically equivalent. Thus setting $r(Q, \hat{Q}, \relB)):= \relB$ yields the desired parsimonious slice-reduction.
\end{proof}

Let $\hat{\boA}^*:=\{(\fadorn(\hat{Q}))\mid \hat{Q} \in \hat{\boA}\}$. Note that, by Lemma~\ref{lem:constants}, there is a counting slice reduction from $\smallp\sCQ[\hat{\boA}^*]$ to $p$-$\sCQ[\hat{\boA}]$.

 Let now $\calG$ be the class of simple queries associated to the queries in $\hat{\boA}$.
 
\begin{claim}\label{clm:graphtohatstar}
 There is a parsimonious slice reduction from $\sCQ[\calG]$ to $\sCQ[\hat{\boA}^*]$.
\end{claim}
\begin{proof}
The relation $U$ relates every simple query $Q_s$ in $\calG$ to all queries $\hat{Q}^*$ such that $Q_s$ is the simple query of $\hat{Q}\in \hat{\boA}$ and $\hat{Q}^*=\fadorn(\hat{Q}$. Certainly, $U$ is computable and by definition of $\calG$ it assigns to every query in $\calG$ a query in $\hat{\boA}^*$.

It remains to describe the function $r$.
So let $(Q_s,\relB)$ be a $p$-$\sCQ$-instance and $\hat{Q}^*$ as above. 
 We assume w.l.o.g.~that all atoms in $Q_s$ that share the same variable scope also have the same relation in $\relB$. If this is not the case, we can modify $\relB$ by taking intersections of some relations in the obvious way.

Let $\tau$ be the vocabulary of $\bar{Q}$.
We construct a structure $r(Q_s, \hat{Q}^*,\relB)=: \hat{\relB}$ over relation symbols of~$\hat{Q}^*$, i.e., over the vocabulary $\tau \cup \{r_X\mid X\in \vars(Q)\}$. The structure $\hat{\relB}$ has the domain $\hat{B}:=\vars(Q_s)\times B$ where $B$ is the domain of $\relB$.
 For $r\in \tau$ we set 
 \begin{align*}r^{\hat{\relB}}:=\{ ((X_1, b_1), \ldots , (X_k, b_k)) \mid & r(X_1, \ldots, X_k)\in r^{\bar{Q}^*}, 
 r'(X_1, \ldots, X_k)\in \atoms(Q_s), \\&(b_1, \ldots, b_k)\in r'^\relB\}.\end{align*}
 Furthermore, for the relations symbols $r_X$ that are added in the construction of $\hat{Q}^*$ from $\hat{Q}$, we set $r_X^{\hat{\relB}}:= \{(X, b)\mid b\in B\}$.
 
It is easy to see that from a satisfying assignment $h:Q_s \rightarrow \relB$ we get a homomorphism $h':Q \rightarrow \hat{\relB}$ by setting $h'(a):= (a, h(a))$. Furthermore, this construction is obviously bijective. Thus we get $|Q_s(\relB)| = |Q(\hat{\relB})|$. 
Since $\hat{\relB}$ can  be constructed in polynomial time in $\|Q_s\|$ and $\|\relB\|$, this is a parsimonious slice reduction.
\end{proof}

\begin{cor}\label{cor:graphstostructures}
Let $\boA$ be a class of conjunctive queries of bounded arity and let~$\calG$ be the simple queries associated to the cores of the colorings of $\boA$. Then there is a counting slice reduction from $\sCQ[\calG]$ to $\sCQ[\boA]$.
\end{cor}
\begin{proof}
 By Claim~\ref{clm:graphtohatstar} there is a reduction from $\sCQ[\calG]$ to $\sCQ[\hat{\boA}^*]$. Moreover, the colorings of the queries in $\hat{\boA}$ are by definition cores, so by Lemma~\ref{lem:constants} there is a reduction from $\sCQ[\hat{\boA}^*]$ to $\sCQ[\hat{\boA}]$. Finally, by Claim~\ref{clm:hattowithouthat} there is a reduction from $\sCQ[\hat{\boA}]$ to $\sCQ[\boA]$. Now using transitivity of counting slice reductions completes the proof.
\end{proof}

\subsection{Frontier Size}

In this section, we consider the influence of the size of frontiers in queries. 
To this end, we define the \emph{frontier size} of a query $Q$ to be the maximum size of any frontier $\Fr(Y,\free(Q),\HG_Q)$, over quantified variables $Y\in\vars(Q)\setminus\free(Q)$.

The aim of this section is the following Lemma.

\begin{lem}\label{lem:strictSS}
 Let $\calG$ be a class of simple queries of bounded arity. If the frontier size of the queries in $\calG$ is unbounded, then there is a counting slice reduction from $\sclique[\mathbb{N}]$ to $\sCQ[\calG]$.
\end{lem}

We start with some easy observations. For a query $Q$ we define the query $\graph(Q)$ to be the simple query that has for every edge $e=uv$ in the primal graph of $Q$ a binary atom $r_e(u,v)$.

\begin{observation}\label{obs:primal}
 Let $\calG$ be a class of simple queries of bounded arity. Define $\calG' := \{ (\graph(Q) \mid Q\in \calG\}$. Then there is a parsimonious slice reduction from $\sCQ[\calG']$ to $\sCQ[\calG]$.
\end{observation}
\begin{proof}
Since the arity of the the queries in  $\calG$ is bounded by a constant $c$, we can for every $\graph(Q)\in \calG'$ simulate the binary relations on the edges of the primal graph of $Q$ by the atoms of $Q$. Note that every atom of $Q$ only has to simulate at most $c^2$ binary relations.
\end{proof}

\begin{observation}\label{obs:subgraph}
 Let $\calG$ be a class of simple queries of bounded arity. Let $\calG'$ be the closure of $\calG$ under deletions of atoms. Then there is a parsimonious slice reduction from $\sCQ[\calG']$ to $\sCQ[\calG]$.
\end{observation}
\begin{proof}
For every atom $r({\bf u})$ not appearing in a query $Q'\in \calG'$ but in $Q\in \calG$, let the respective relation $r^\DB$ in the database contain all $k$-tuples of domain elements, where $k$ is the arity of $r$. Since $k$ is bounded by a constant, this database can be constructed in polynomial time.
\end{proof}

The following lemma is an easy translation of a result from \cite{DurandM15} into our framework.

\begin{lem}[\cite{DurandM15}]\label{lem:starsize}
 Let $\calG$ be a class of simple queries of unbounded quantified star size. Then there is a counting slice reduction from $\sclique[\mathbb{N}]$ to $\sCQ[\calG]$.
\end{lem}

We now have everything in place to prove the main result of this section.

\begin{proof}[Proof of Lemma~\ref{lem:strictSS}]
 For every simple query $Q$ we compute a simple query as follows: Take the query $\graph(Q)$ and delete all atoms containing only free variables. Let $\calG'$ be the resulting class of simple queries. Obviously, $\calG'$  has unbounded quantified star size and thus there is a counting slice reduction from $\sclique[\mathbb{N}]$ to $\sCQ[\calG']$. Using Observation \ref{obs:subgraph} and Observation \ref{obs:primal} then gives the desired result.
\end{proof}

\subsection{The main hardness results}

In this section we use the results of the last sections to prove the hardness results of Lemma~\ref{lem:hardness}.
Recall that we are focusing here on bounded-arity relational structures, 
 for which it is well known that a class of hypergraphs have bounded (generalized) hypertree width if and only if they have bounded treewidth (more precisely, their associated primal graphs do). 
 Thus, we use the two notions interchangeably in this section, 
 in order to simplify the references to previous results on treewidth described in the literature.

We take the following result from~\cite{DurandM15}.

\begin{lem}\label{lem:DMreduction}
 Let $\calG$ be a class of simple queries of bounded arity. If the treewidth of~$\calG$ is unbounded, then there is a counting slice reduction from to $\pn{Clique}[\mathbb{N}]$ to $\sCQ[\calG]$.
\end{lem}

Combining Lemma~\ref{lem:DMreduction} and Corollary~\ref{cor:graphstostructures} yields the following corollary.
\begin{cor}\label{cor:decision}
 Let $\boA$ be a class of queries of bounded arity such that the treewidth of the cores of the queries in $\boA$ is unbounded. Then there is a counting slice reduction from $\clique[\mathbb{N}]$ to $\sCQ[\boA]$.
\end{cor}
\begin{proof}
 As in Corollary~\ref{cor:graphstostructures}, let $\calG$ be the class of simple queries associated to the cores of the queries in the class $\boA$. By assumption, $\calG$ is of unbounded treewidth and thus there is a counting slice reduction from $\clique[\mathbb{N}]$ to $\sCQ[\calG]$. But by Corollary~\ref{cor:graphstostructures} we have a counting slice reduction from $\sCQ[\calG]$ to $\sCQ[\boA]$. Consequently, there is a counting slice reduction from $\clique[\mathbb{N}]$ to $\sCQ[\boA]$.
\end{proof}

To deal with the case of unbounded treewidth, we use the following result by Dalmau and Jonsson.

 \begin{thm}[\cite{DJ04}]\label{thm:dalmauJ}
Let $\calQ$ be the class of all quantifier free conjunctive queries. Let $\boA$ be a recursively enumerable class of queries of bounded arity in $\calQ$. Assume $\FPT \ne \swone$. Then the following statements are equivalent:
\begin{enumerate}
 \item $\lpr \sCQ, \boA \times \Sigma^* \rpr \in \prom{\Pol}$.
 \item $\lpr \smallp\sCQ, \boA \times \Sigma^* \rpr \in \prom{\FPT}$.
 \item There is a constant $c$ such that the queries in $\boA$ have treewidth at most~$c$.
\end{enumerate}
\end{thm}

We get the following hardness result.
\begin{lem}\label{lem:contracthard}
 Let $\calG$ be a class of simple queries of bounded arity whose frontier hypergraphs $\Frontiers(\calG)$ have unbounded hypertree width. Then there is a counting slice reduction from $\sclique[\mathbb{N}]$ to $\sCQ[\calG]$.
\end{lem}
\begin{proof}
 Assume first that $\calG$ is of unbounded frontier size. Then the claim follows by Lemma~\ref{lem:strictSS}. So we assume in the remainder of the proof that there is a constant~$c$ such that for every $Q\in \calG$ every frontier $\Fr(Y,\free(Q))$ of $Q$ contains only $c$ vertices from~$\free(Q)$.
 
 To every frontier hypergraph  $\HG=\FH(Q, \free(Q))$, we assign a quantifier free, simple query $Q'$ by introducing for every edge $e$ in $\HG$ an atom that contains exactly the variables in $e$. Note that the hypergraph $\HG_{Q'}$ is exactly $\HG$. Let $\calG'$ be the class of simple queries constructed this way.
 
 Since the hypertree width (and hence the treewidth) of the frontier hypergraphs of queries in $\calG$ is unbounded, 
 and thus also the treewidth of the queries in $\calG'$ is unbounded, it follows from Theorem~\ref{thm:dalmauJ} that there is a counting slice reduction from $\sclique[\mathbb{N}]$ to $\sCQ[\calG']$. Therefore, it suffices to show a parsimonious slice reduction $(U,r)$ from $\sCQ[\calG']$ to $\sCQ[\calG]$ to show the lemma.
 
 The relation $U$ is defined as $U:=\{(Q', Q)\mid Q \in \calG\}$. By definition, this satisfies the covering condition.
 
For the definition of $r$, consider an instance $(Q', \relB')$ of $\sCQ[\calG']$ and let $\HG$ be the hypergraph of $Q'$. Remember that, by construction, for every edge $e$ of the frontier graph $\Fr(Q, \free(Q))$ the query $Q'$ contains an atom with the variables in $e$ in some order. We construct a structure $r(Q', Q, \relB')) = \relB$. For every $\component{\free(Q)}$ $C$ of $Q$ we do the following: Let $D_C$ be the tuples encoding the homomorphisms $h$ from $Q[\Fr(Y,\free(Q), \HG_{Q})]$ to $\relB$ for a variable in $C$. For every variable $Y$ of $C$ we let $D_C$ be the domain of $Y$. Whenever two elements $Y,Y' \in V(C)$ appear as variables in an atom with relation symbol $r$, we set $r^{\relB}$ in such a way that for all tuples the assignments to $Y$ and $Y'$ coincide. Moreover, whenever $X\notin V(C)$ and $Y\in V(C)\setminus S$, i.e., $X\in \Fr(Y,\free(Q), \HG_{Q})$, we allow only tuples in which the assignment to $X$ coincides with the assignment to $X$ that is encoded in the assignment to $Y$. For all atoms $r(\bar{u})$ of $Q'$ with $\bar{u}\setminus \free(Q)\ne \emptyset$, we let $r^{\relB}$ contain all tuples that satisfy the two conditions. Finally, for all atoms that have only free variables, we set $r^{\relB} :=r^{\relB'}$.

It is easy to verify that $Q(\relB) = Q'(\relB')$. Thus it only remains to show that the construction can be done in polynomial time. Note first that the number of free variables in any $\Fr(Y,\free(Q), \HG_{Q})$ is bounded by $c$. Thus we can compute all domains $D_C$ in time $\|\relB\|^{O(c)}$. The rest of the construction can then be easily done in polynomial time.
\end{proof}


%% file: appendixNew.tex

\begin{appendix}

\section{On the relationship between Quantified Starsize and {\sf \#}-{hypertree width}}\label{appendix:starsize}

In this section, we contrast {\sf \#}-hypertree decompositions with the approach to counting answers of conjunctive queries with existential variables that was independently proposed by Durand and Mengel in~\cite{DurandM15}.
It is founded on the concept of \emph{quantified star size}, which can be recast, in our notation and terminology, as
the cardinality of the maximum independent set over the frontiers $\Fr(Y,\free(Q),\HG_Q)$, for any quantified variable $Y$, i.e.,
any $Y\in \vars(Q)\setminus\free(Q)$.

\begin{prop}[cf.\cite{DurandM15}]\label{thm:durandt}
Let $k$ and $\ell$ be fixed natural numbers. Given any query $Q$ having quantified star size at most $\ell$, together with a generalized
hypertree decomposition of $Q$ of width at most $k$, $\countProblem(Q,\DB)$ can be solved in polynomial time, for every database $\DB$.
\end{prop}

Contrast Theorem~\ref{thm:general} and Definition~\ref{def:main} with Theorem~\ref{thm:fptGHTW}.
It can be immediately recognised that cores are not considered in the notion defined by~\cite{DurandM15}, so that it is easy to see that
there are classes of tractable instances that are not recognised according to this notion, but that are tractable according to
 the notion of {\sf \#}-hypertree width.
 
\begin{figure}[h]
 \centering
    \includegraphics[width=0.55\textwidth]{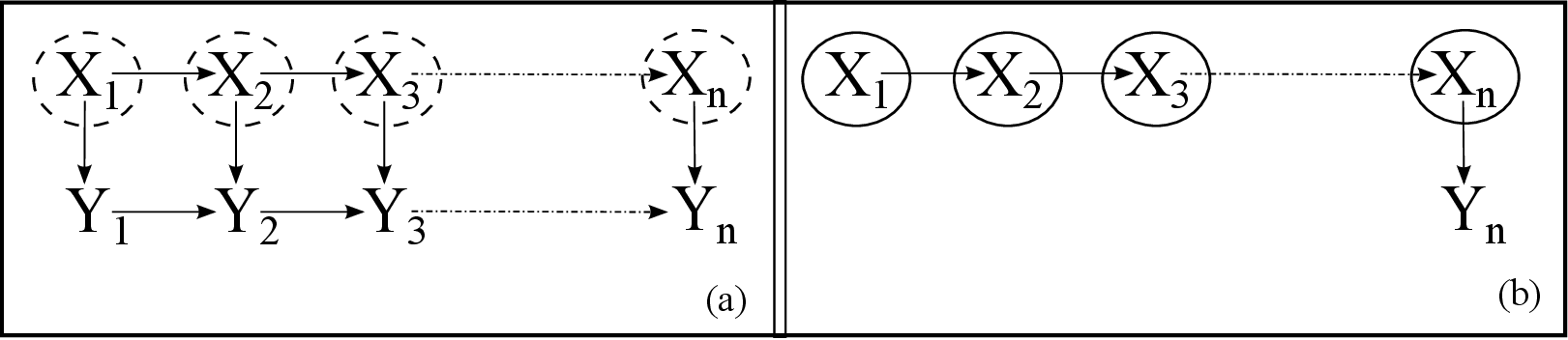}
  \caption{The hypergraphs $\HG_{Q^n_1}$ and $\HG_{\bar \Phi^n_1}$ in Example~\ref{ex:esempio3}. Edge orientation reflects the position of the variables in the relation $r$.}\label{fig:esempio3}
\end{figure}

\begin{exa}\label{ex:esempio3}
Consider the class of queries $\{ Q^n_1= \exists Y_1,...,Y_n \Phi_n \mid n>0\}$, with $\free(Q^n_1)=\{X_1,...,X_n\}$ and
$$
\hspace{-1mm}
\begin{array}{l}
   \Phi_n =  \bigwedge_{i=1}^{n} r(X_{i},Y_{i})\wedge \bigwedge_{i=1}^{n-1} r(X_{i},X_{i+1})\wedge \bigwedge_{i=1}^{n-1} r(Y_{i},Y_{i+1}).\\

\end{array}
$$

The (hyper)graph associated with such a query $Q^n_1$ is shown in Figure~\ref{fig:esempio3}(a), 
where dashed hyperedges correspond to the free variables.
Every query in the class has hypertree width 2. However, the quantified star size is $\lceil n/2 \rceil$, because all the free
variables $X_1,...,X_n$ belong to the frontier of, e.g., the quantified variable $Y_1$. Hence, this class is not recognized as tractable
according to Proposition~\ref{thm:durandt}.

On the other hand, 
note that the acyclic (sub)query $\bar \Phi^n_1= r(X_{n},Y_{n})\wedge \bigwedge_{i=1}^{n-1} r(X_{i},X_{i+1})\wedge\bigwedge_{i=1}^n
r_{X_i}(X_i)$ is the core of $\adorn(Q^n_1)$, where each variable $Y_i$, with $i\in\{1,...,n-1\}$, is mapped to $X_{i+1}$. Its associated
hypergraph is shown in Figure~\ref{fig:esempio3}(b). There, check that $Y_n$ is the only quantified variable and that its frontier is the
singleton $\{X_n\}$. It follows that, for every $n > 0$, $Q^n_1$ has {\sf \#}-{hypertree width} $1$,
which entails that this class of queries is tractable, by Theorem~\ref{thm:fptGHTW}. 
\hfill
$\lhd$
\end{exa}
 
We say that a class of queries $\boA$ has {\em bounded hypertree width and bounded quantified star size} if there are two finite natural
numbers $k$ and $\ell$ such that each query $Q\in\boA$ has quantified star size at most $\ell$, and there exists a generalized hypertree
decomposition of $Q$ of width at most $k$.


Example~\ref{ex:esempio3}  shows a class of {\rm FPT} instances having unbounded
quantified star size. In fact, it turns out that the notion of {\sf \#}-{hypertree width} allows us to identify larger classes of
fixed-parameter tractable queries, and in this paper we have shown that it precisely charts the tractability frontier.

\begin{thm}\label{thm:comparison}
Any class of queries having {bounded hypertree width and bounded quantified star size} has bounded {\sf \#}-{hypertree width}, too.
However, there are classes having bounded {\sf \#}-{hypertree width}, but unbounded hypertree width or unbounded quantified star
size.
\end{thm}
\begin{proof}
The latter statement immediately follows from Example~\ref{ex:esempio3}. Indeed, the class of queries $\{ Q^n_1 \mid n>0 \}$ described in this
example has unbounded quantified starsize but bounded {\sf \#}-{hypertree width}. For completeness, we point out that there are also classes 
of queries having unbounded hypertree width, but bounded {\sf \#}-{hypertree width}: 
for example, consider $\{ Q^n_2= \exists Y_1,...,Y_n \Phi_n \mid n>0\}$, with
$$
\hspace{-1mm}
\begin{array}{l}
   \Phi_n =  \bigwedge_{i,j=1}^{n} r(X_{i},Y_{j})\\

\end{array}
$$
and with $\free(Q^n_2)=\emptyset$, i.e., all variables a projected away. Then $\adorn(Q_2^n) = Q_2^n$ and the core of $\adorn(Q_2^n)$ is $\bar \Phi^n_1= r(X_{1},Y_{1})$, so  
$Q^n_1$ is \sharpcovered\ w.r.t.~$\V^1_{Q_1^n}$. Thus the {\sf \#}-{hypertree width} of $Q_2^n$ is bounded but it is easy to see that the (generalized) hypertree width of $Q_2^n$ is $n$.

Consider now any class of queries $\boA$ having bounded hypertree width and bounded quantified starsize, with $k'$ and $\ell'$ being the bounds
for the generalized hypertree width and the quantified star size of the queries in $\boA$, respectively. Let $Q$ be any query in $\boA$, and
let $\HG_a$ be a width-$k$ generalized hypertree decomposition for it, with $k\leq k'$. Let $Q'$ be any core of $\adorn(Q)$, and note that
$\HG_a$ is also a tree projection of $\HG_{Q'}$ with respect to the set of views $\VQ^{k}$, in particular $\HG_{Q'}\leq \HG_a$ holds. Let us
now build a hypergraph  $\HG_a'$ from $\HG_a$ as follows. For each hyperedge $h\in \edges(\HG_a)$, we add to $\HG_a$ the hyperedge $h'=h\cap
\free(Q) \cup \bigcup_{Y\in h\setminus\free(Q)} \Fr(Y,\free(Q),\HG_Q). $ By construction, $\HG_a'$ covers the hypergraph $\FH(Q,\free(Q))$, and
thus also the hypergraph $\FH(Q',\free(Q))$, because it is easy to see that the frontier of every quantified variable can only be smaller if
the core $Q'$ is considered, instead of the full query $Q$. Thus,  we have $\HG_{Q'}\leq \HG_a\leq \HG_a'$ and $\FH(Q',\free(Q))\leq
\FH(Q,\free(Q))\leq \HG_a'$. Moreover, since each hyperedge $h\in \edges(\HG_a)$ covers the variables contained in at most $k$ query atoms in
$Q$, $h$ can intersect at most $k$ distinct hyperedges of $\FH(Q,\free(Q))$ (by definition of frontiers, every atom can deal with one frontier
at most, because its existential variables belong to the same connected component). If the quantified starsize of $Q$ is $\ell\leq \ell'$,
$\Fr(Y,\free(Q),\HG_Q)$ is covered by at most $\ell$ hyperedges of $\HG_a$ for each $Y\in \vars(Q)\setminus\free(Q)$. So, $\HG_{a}'$ is covered
by the hypergraph associated with $\VQ^{k\times \ell}$. Moreover, it can be seen that $\HG_a'$ is acyclic, and hence it is a width-$(k\times
\ell)$ \sharpHD\ of $Q$.
\end{proof}

Analyzing Example~\ref{ex:esempio3}, one sees that the quantified star size of the core of $Q^n$ is $1$. 
Thus, when taking the core before 
computing the quantified star size, the separation between quantified star size and {\sf \#}-{hypertree width} goes away on the 
example. We next show that in fact this is a general phenomenon.

\begin{lem}\label{lem:comparison2}
 Let $\boA$ be query having bounded {\sf \#}-{hypertree width} $k$. Then the cores of the colorings of the queries in
 $\boA$ have bounded  hypertree width and bounded quantified star size at most $k$.
\end{lem}
\begin{proof}
 Note first, that by definition any ${\sf \#}$-hypertree decomposition of $Q'$ is also a generalized hypertree decomposition of $Q'$. Consequently,
 the {\sf \#}-{hypertree width} is at least the generalized hypertree width of $Q'$.
 
 Now assume that the quantified star size of $Q'$ is at least $k+1$. Then there is by definition a frontier $\Fr(Y,\free(Q'))$ for some quantified 
 variable $Y$ that contains variables $Y_1, \ldots, Y_{k+1}$ that form an independent set of $\HG_{Q'}$. Thus, to cover $\Fr(Y,\free(Q'))$ in a 
 ${\sf \#}$-hypertree decomposition one needs at least $k+1$ edges of $\HG_{Q'}$. Thus, the 
 {\sf \#}-{hypertree width} must be at least $k+1$ which contradicts the assumption.
\end{proof}

From Theorem~\ref{thm:comparison} and Lemma~\ref{lem:comparison2}, we directly get the following corollary.

\begin{cor}
 A class $\calC$ of queries has bounded {\sf \#}-{generalized hypertree width} if and only if the generalized hypertree width and the quantified
 star size of the cores of the colorings of queries in $\calC$ is bounded.
\end{cor}

\section{Proofs in the Case Complexity Section} 
\label{app:case}

\begin{proof}[Proof of Theorem~\ref{thm:transitivity}]
Suppose that $(U_1, r_1)$ is a counting slice reduction from
$Q_1[S_1]$ to $Q_2[S_2]$, 
and that $(U_2, r_2)$ is a counting slice reduction from
$Q_2[S_2]$ to $Q_3[S_3]$.
We show that there exists a counting slice reduction
$(U, r)$ from $Q_1[S_1]$ to $Q_3[S_3]$.

Define $U \subseteq \Sigma^* \times \powfin(\Sigma^*)$
to be the set that contains a pair $(s_1, T_3)$
if and only if
there exists $T_2$ such that $(s_1, T_2) \in U_1$;
for each $t_2 \in T_2$, there exists
$T_{t_2}$ such that $(t_2, T_2) \in U_2$;
and,
it holds that $\bigcup_{t_2 \in T_2} T_{t_2} = T_3$.

We verify the coverage condition as follows.
For each $s_1 \in S_1$, there exists $T_2 \subseteq S_2$
such that $(s_1, T_2) \in U_1$ (since coverage holds for $U_1$),
and for each $t_2 \in T_2$,
there exists $T_{t_2}$ such that $(t_2, T_{t_2}) \in U_2$
(since coverage holds for $U_2$).
Hence, by definition of $U$, it holds that
$(s, \bigcup_{t_2 \in T_2} T_2) \in U$.

We verify the correctness condition as follows.
Let $A_1$ and $A_2$ by the algorithms given by the definition
of counting slice reduction for $r_1$ and $r_2$, respectively.
We describe an algorithm $A$ for the needed partial function $r$,
as follows; the algorithm $A$ uses the algorithms 
$A_1$ and $A_2$ in the natural fashion.
On an input $(s_1, T_3, x)$, the algorithm $A$
checks if $(s_1, T_3) \in U$; if so,
it may compute a set $T_2$ and sets $\{ T_{t_2} \}_{t_2 \in T_2}$
that witness this (in the sense of the definition of $U$).
The algorithm $A$ then invokes the algorithm $A_1$
on input $(s_1, T_2, x)$; each time $A_1$
makes an oracle query,
it is of the form $Q_2(t_2, z)$ where $t_2 \in T_2$,
and to resolve the query, the algorithm $A$ calls
$A_2$ on input $(t_2, T_{t_2}, z)$.

The time analysis of the algorithm $A$ is analogous to
that carried out in~\cite[Appendix Section A]{Chen14}.
\end{proof}

 \begin{proof}[Proof of Theorem~\ref{thm:slice-red-gives-fpt-red}]
Let $(U, r)$ be the counting slice reduction.
Since $S$ and $S'$ are both computable,
there exists a computable function $h: \Sigma^* \to \powfin(\Sigma^*)$
such that for each $s \in S$, it holds that $(s, h(s)) \in U$.
Consider the algorithm $A'$ that does the following:
given $(s, y)$, check if $s \in S$; if not, then output $0$,
else compute $r(s, h(s), y)$ using the algorithm $A$ for $r$
guaranteed by the definition of counting slice reduction.
This algorithm $A'$ and $h$ give a counting FPT-reduction
from $\param{Q[S]}$ to $\param{Q'[S']}$.
Note that, as a function of $(s,y)$,
we have that $A'(s,y)$ is FPT-computable with respect to $\pi_1$,
since $r(s,h(s),y)$ is FPT-computable with respect to $(\pi_1,
\pi_2)$,
and $h(s)$ is a function of $s$.
\end{proof}

\begin{proof}[Proof of Theorem~\ref{thm:promfpt-to-fpt}]
Let $(U, r)$ be the counting slice reduction given by hypothesis.
Since $U$ and $S$ are both computably enumerable,
there exists
 a computable function $h: \Sigma^* \to \powfin(\Sigma^*)$
such that for each $s \in \Sigma^*$, it holds that $(s, h(s)) \in U$.
Consider the following algorithm $A'$ for $K$:
on an input $(s, y)$,
 compute $r(s, h(s), y)$ using the algorithm $A$ for $r$
guaranteed by the definition of counting slice reduction;
the oracle queries that the algorithm $A$ poses to $Q$ are resolved
by the algorithm $B$ which witnesses $\paramprom{Q[S]} \in \prom{\FPT}$.
Suppose that the running time of $B$ on $(t, z)$ is bounded above by
$f(t)p(|(t,z)|)$; note that the time needed to resolve an oracle query
$(t,z)$
made by $A$ on $(s,y)$ is
$(\max_{t \in h(s)} f(t)) p(|(t,z)|)$, and that 
$\max_{t \in h(s)} f(t)$ is a computable function of $s$.
We then have that $A'(s, y)$ is FPT-computable with respect to $\pi_1$,
by an argument similar to that in the proof of
Theorem~\ref{thm:slice-red-gives-fpt-red}.
\end{proof}

\section{An Algorithm for ``Bounded-degree'' Databases}\label{sec:bounded}

Pichler and Skritek~\cite{PS13} have proposed an algorithm for counting answers to acyclic conjunctive queries, whose scaling is in
the worst case exponential w.r.t.~the maximum number of tuples over the database relations,  denoted by $m$ hereinafter. In fact, they also
shown that counting query answers remains $\rm \#P$-hard over this structurally simple class (of course, they considered acyclic queries where existential variables are not ``guarded'', by considering their frontier hypergraphs).
Moreover, they pointed out that their algorithm can be extended to queries that are not necessarily acyclic, but have bounded \emph{hypertree
width}~\cite{gott-etal-99}. Our first step towards proposing a hybrid decomposition method for counting problems is to analyze such an
extension.

\begin{figure}[t]
 \centering
    \includegraphics[width=0.55\textwidth]{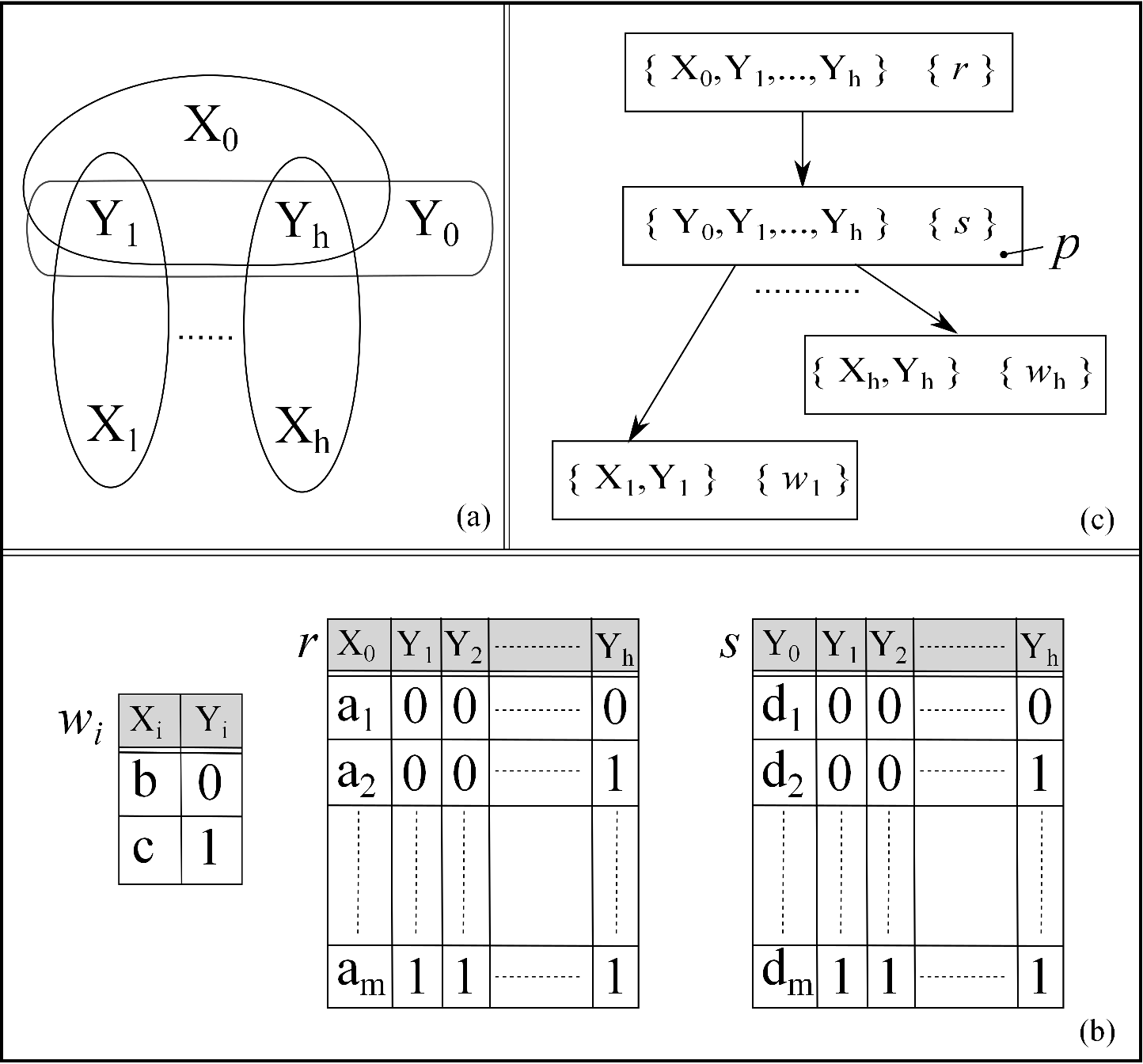}
  \caption{Structures in Example~\ref{ex:hybrid1}: (a) The hypergraph $\HG_{Q^h_2}$; (b) The database $\DB_2$; (c) A hypertree decomposition $\HD_2$ for the query ${Q^h_2}$.}\label{fig:hybrid1}
\end{figure}

\smallskip

A \emph{hypertree decomposition}~\cite{gott-etal-99} of $Q$ is a generalized hypertree decomposition enjoying an additional property, often
called {\em descendant condition}. It is now convenient to recall its direct definition (instead of using a view set for $Q$). Formally, a {\em
hypertree} for a query $Q$ is a triple $\tuple{T,\chi,\lambda}$, where $T$ is a rooted tree, and $\chi$ and $\lambda$ are labeling functions
associating each vertex $p$ of $T$ with two sets $\chi(p)\subseteq \vars(Q)$ and $\lambda(p)\subseteq \atoms(Q)$.
The set of the vertices of $T$ is denoted by $\vertices(T)$, whereas its root is denoted by $\root(T)$.
Moreover, for any $p\in N$, we denote by $T_p$ the subtree of $T$ rooted at $p$ (hence, $T=T_{\root(T)})$, and by $\chi(T_p)$ the set of all
variables occurring in the $\chi$ labeling of $T_p$.

A \emph{hypertree decomposition} of $Q$ is a hypertree $\HD=\tuple{T,\chi,\lambda}$ for $Q$ such that: (1) for each $q\in \atoms(Q)$, there
exists $p\in \vertices(T)$ such that $\vars(q)\subseteq \chi(p)$; (2) for each $X\in \vars(Q)$, the set $\{ p\in \vertices(T) \mid X \in
\chi(p) \}$ induces a (connected) subtree of $T$; (3) for each $p\in \vertices(T)$, $\chi(p)\subseteq \vars(\lambda(p))$; and (4) for each
$p\in \vertices(T)$, $\vars(\lambda(p)) \cap \chi(T_p) \;\subseteq\; \chi(p)$ (descendant condition).
The decomposition $\HD$ is said \emph{complete} if for each atom $q\in \atoms(Q)$, there exists $p\in \vertices(T)$ such that $q\in
\lambda(p)$.

This notion is a true generalization of acyclicity, as acyclic queries are precisely those having hypertree width 1.
Importantly, for any fixed natural number $k\geq 1$, deciding whether a query has hypertree width at most $k$ is in $\LCFL$, and thus it is a
tractable and highly parallelizable problem~\cite{gott-etal-99}.

\begin{exa}\label{ex:hybrid1}
For\hspace{-0.5mm} any\hspace{-0.5mm} natural\hspace{-0.5mm} number $h\geq 1$,\hspace{-0.5mm} consider the query 
%
$$Q_2^{h}\ =\ \exists Y_0,...,Y_h\ r(X_0,Y_1,...,Y_h)\wedge s(Y_0,Y_1,...,Y_h)\ \wedge 
                       \bigwedge_{i\in\{1,...,h\}} w_i(X_i,Y_i),$$
whose hypergraph is in Figure~\ref{fig:hybrid1}(a).

Observe that the frontier of any existentially quantified variable is the whole set $\{X_0,X_1,...,X_h\}$ of the free variables. Hence, $Q_2^{h}$
is not \sharpcovered\ w.r.t.~$\VQ^{k}$, for each $k<h+1$.

However, it is immediate to check that $Q^h_2$ is acyclic. This is witnessed by the complete hypertree decomposition $\HD_2$ for $Q_2^h$,
reported in Figure~\ref{fig:hybrid1}(c) and whose width is 1.\hfill $\lhd$
\end{exa}

\smallskip

\newcounter{stepcount}
\newcommand{\step}{%
        \stepcounter{stepcount}%
        \thestepcount.}
\newcommand{\resetstep}{%
        \setcounter{stepcount}{0} } \renewcommand{\step}{\ \ \, \ }
\begin{figure}[t]
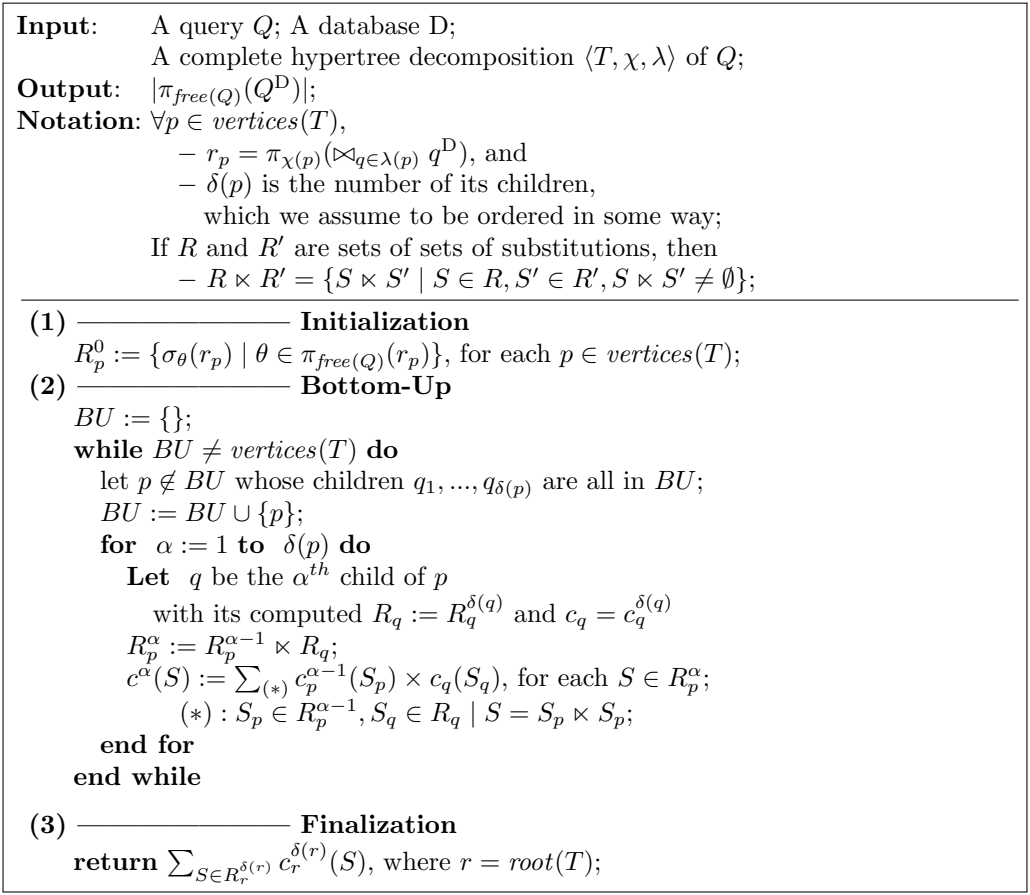

\centering
\fbox{
\parbox{0.87\textwidth}
{\small
\begin{tabular}{@{\hspace{-0.6mm}}l@{\hspace{0.0mm}}l}
\textbf{Input}:     &\ A query $Q$; A database $\DB$; \\
                    &\ A complete hypertree decomposition $\tuple{T,\chi,\lambda}$ of $Q$;\\
\textbf{Output}:    &\ $|\pi_{\free(Q)}(Q^\onDB)|$;\\
\textbf{Notation}:  &\ $\forall p\in \vertices(T)$,\\
                    &\ \quad $-$ $r_p=\pi_{\chi(p)} (\bowtie_{q\in \lambda(p)} q^\onDB)$, and\\ 
                    &\ \quad $-$  $\delta(p)$ is the number of its children,\\
                     &\ \quad \quad which we assume to be ordered in some way;\\
                    &\ If $R$ and $R'$ are sets of sets of substitutions, then    \\
                    &\ \quad $-$ $R \ltimes R'=\{S\ltimes S' \mid S\in R, S'\in R', S\ltimes S'\neq \emptyset\}$;
 \end{tabular} \\
\hrule\vspace{1mm}
\hspace{-1mm}
\begin{tabular}{@{\hspace{-0.3mm}}r@{\hspace{3mm}}l@{\hspace{0.9mm}}l}
            & \hspace{-6mm}\emph{\bf (1) ---------------------  Initialization}\\
   \step    & $R^0_p:=\{ \sigma_\theta(r_p)\mid \theta\in \pi_{\free(Q)}(r_p)\}$,  for each $p\in \vertices(T)$;\\
            & \hspace{-6mm}\emph{\bf (2) ---------------------  Bottom-Up}\\
   \step    & $BU:=\{ \}$;\\
   \step    & \textbf{while} $BU\neq \vertices(T)$ \textbf{do}\\
   \step    & \quad let $p\not\in BU$ whose children $q_1,...,q_{\delta(p)}$ are all in $BU$;\\
   \step    & \quad $BU:=BU\cup \{p\}$; \\
   \step    & \quad \textbf{for } $\alpha := 1$ \textbf{to } $\delta(p)$  \textbf{do}\\
   \step    & \quad \quad \textbf{Let } $q$ be the $\alpha^{th}$ child of $p$  \\
   \step    & \quad \quad \quad with its computed $R_q:=R_q^{\delta(q)}$ and $c_q=c^{\delta(q)}_q$\\
   \step    & \quad \quad $R^\alpha_p:=R^{\alpha-1}_p\ltimes R_q$;\\
   \step    & \quad \quad $c^\alpha(S):=\sum_{(*)}c^{\alpha-1}_p(S_p)\times c_q(S_q)$, for each $S\in R^\alpha_p$;\\
   \step    & \quad \quad \quad \quad $\scriptsize {(*)}: S_p\in R^{\alpha-1}_p, S_q\in R_q \mid S=S_p\ltimes S_p$;\\
   \step    & \quad \textbf{end for}\\
   \step    & \textbf{end while}\\  \vspace{-2mm} \ \\
            & \hspace{-6mm}\emph{\bf (3) --------------------- Finalization}\\
   \step    & \textbf{return} $\sum_{S\in R^{\delta(r)}_{r}} c^{\delta(r)}_{r}(S)$, where $r=\root(T)$;\\
\end{tabular}
}}\vspace{-2mm}
\caption{Counting query answers via hypertree decompositions.}\label{fig:algoritmo}
\end{figure}

With the above notation in place, consider the algorithm in Figure~\ref{fig:algoritmo}, which takes as input a complete hypertree decomposition
$\HD=\tuple{T,\chi,\lambda}$ of a query $Q$, and a database $\DB$. It is articulated in three phases.
In the initializations step, for each vertex $p$, the set $r_p$ (of the partial solutions over $p$) is partitioned according to the profiles
given by the configurations of values allowed for the free variables in $\chi(p)$. Let $R^0_p$ be the resulting partition.
In the subsequent steps, $p$ will be associated with a set $R^\alpha_p$ where $\alpha\in\{0,1,...,\delta(p)\}$, with $\delta(p)$ being the
number of children of $p$. In particular, each of these sets consists of a set of substitutions that are manipulated via an ad-hoc semijoin
``$\ltimes$'' operator such that: if $R$ and $R'$ are sets of sets of substitutions (hereinafter called \emph{{\sf \#}-relations}), then  $R
\ltimes R'=\{S\ltimes S' \mid S\in R, S'\in R', S\ltimes S'\neq \emptyset\}$.
In the bottom-up evaluation, we perform upward semijoins, and $R^\alpha_p$ is the {\sf \#}-relation associated with $p$ after that the first
$\alpha$ children (according to some fixed ordering) have been processed.
Each set $S\in R^\alpha_p$ is associated with a value $c^\alpha_p(S)$ initialized to 1 and updated  bottom-up. At the end, the value computed
in the finalization phase is returned as the desired number of solutions.
The correctness easily follows from the arguments in \cite{PS13}.
In fact, because the running time of the algorithm in~\cite{PS13} is in general exponential w.r.t.~$m$, 
we use the notion of database degree to keep under control such a blow-up.
Intuitively, the {degree} value $\mathit{bound}(\DB,\HD)$ for $Q$ provides an estimate on the size of the information that is required to flow along the given hypertree decomposition $\HD$ in order to answer a counting problem over $\DB$.

\thmmainHybrid*

\begin{proof}
Consider the algorithm in Figure~\ref{fig:algoritmo}, by noticing that it requires as input a hypertree decomposition that is complete. If
$\HD$ is not complete, then we can proceed as follows. For each atom $q\in \atoms(Q)$ such that there is no vertex $p\in \vertices (T)$ with
$q\in \lambda(p)$, let $p_q$ be any vertex such that $\vars(q)\subseteq \chi(p_q)$. Note that $p_q$ is well-defined by condition (1) of
hypertree decompositions. Then, let us modify $\HD$ by adding a fresh vertex $c_q$ as a child of $p_q$ and by setting $\chi(c_q)=\vars(q)$ and
$\lambda(c_q)=\{q\}$. Note that after the modification, $\HD$ is still a hypertree decomposition of $Q$. Moreover, let us update the database
$\DB$ by replacing $q^\onDB$ with $q^\onDB\ltimes (\pi_{\chi(p_q)} (\bowtie_{\bar q\in \lambda(p_q)} \bar q^\onDB))$, i.e., by filtering
$q^\onDB$ deleting those tuples that do not match with its parent $p_q$ (and thus cannot occur in any query answer). Then, the bounded-degree
property over $c_q$ follows from that of its parent. Hence,$\mathit{bound}(\DB,\HD)\leq h$ holds on the resulting database.

We now run the algorithm over the complete decomposition $\HD$.
We claim that, for each vertex $p$ and for each $\alpha\in\{0,1,...,\delta(p)\}$, it holds that: $|R^\alpha_p|\leq m^k\times 2^h$; $|S|\leq h$,
for each $S\in R^\alpha_p$; and $|\{ \pi_{\free(Q)}(S) \mid S\in R^\alpha_p\}|\leq m^k$. The claim is proven by structural induction on
$\alpha\in\{0,1,...,\delta(p)\}$.

The base case is for $\alpha=0$. Observe that, for each $\theta\in \pi_{\free(Q)}(r_p)$, the bound on the degree entails $|\sigma_\theta(r_p)|\leq
h$. Moreover, by definition of $r_p$, $|\pi_{\free(Q)}(r_p)|\leq m^k$ holds. By looking at the initialization step, we then conclude:
$|R^0_p|\leq m^k$; $|S|\leq h$, for each $S\in R^0_p$; and $|\{ \pi_{\free(Q)}(S) \mid S\in R^\alpha_p\}|\leq m^k$.

Assume that the property holds up to any $0\leq \beta < \delta(p)$. We have to show that it holds on $\beta+1$, too.

In fact, in the bottom-up phase, the algorithm defines $R^{\beta+1}_p=R^{\beta}_p\ltimes R_q^{\delta(q)}$ and we are assuming that the claim
holds on $R^{\beta}_p$.
In particular, by inductive hypothesis, the elements of $R^{\beta}_p$ can be partitioned in the sets $C_1^\beta,...,C_\ell^\beta$, with
$\ell\leq m^k$, such that for each $i\in\{1,...,\ell\}$, and for each pair of sets $S$ and $S'$ in $C_i$, it holds that
$\pi_{\free(Q)}(S)=\pi_{\free(Q)}(S')$. By performing semijoin operations, this property is clearly preserved, i.e., we have $|\{
\pi_{\free(Q)}(S) \mid S\in R^{\beta}_p\}|\leq m^k$.
Moreover, for each $i\in\{1,...,\ell\}$ and for each $S\in C_i^\beta$, $|S|\leq h$ holds. Therefore, each of the elements in $C_i^\beta$ can
give rise to at most $2^h$ different subsets as a result of the semijoin operation (no matter of how many elements are in $R_q^{\delta(q)}$),
each again having size bounded by $h$. Thus, we have also derived $|R^{\beta+1}_p|\leq m^k\times 2^h$, and $|S|\leq h$, for each $S\in
R^{\beta+1}_p$.

Armed with the above property, it is now immediate to check that each semijoin operation over {\sf \#}-relations is feasible (with a rough
upper bound) in $O(m^{2\times k}\times 4^h)$. This is the dominant operation and it has to be repeated $|\vertices(T)|$ times in the bottom-up
phase.
\end{proof}

\begin{exa}\label{ex:hybrid2}
Consider the query $Q_2^h$ in Example~\ref{ex:hybrid1}, the database $\DB_2$ in Figure~\ref{fig:hybrid1}(b),
and the hypertree decomposition $\HD_2$ in Figure~\ref{fig:hybrid1}(c). The vertex $p$ in the hypertree does not cover any free variable.
Hence, $\pi_{\free(Q)}(r_v)$ contains only the trivial substitution $\phi$ with empty domain so that $\pi_{\chi(p)}(\sigma_{\phi} (r_p))=m$,
with $m=2^h$, witnessing that $\mathit{bound}(\DB_2,\HD_2)=m$.
In fact, it can be checked that there is no width-$1$ hypertree decomposition $\HD$ such that $\mathit{bound}(\DB_2,\HD)<m$, because of the
relation $s$.

However, even if $\mathit{bound}(\DB_2,\HD_2)=m$, this instance is not that hard. Indeed, consider the hypertree decomposition $\HD_2'$
obtained from $\HD_2$, by ``merging'' the root with its child into a vertex $p'$ such that $\chi(p')=\{X_0,Y_0,Y_1,...,Y_h\}$ and
$\lambda(p')=\{r,s\}$.
In this modified scenario, $\pi_{\free(Q)}(r_{p'})$ consists of the set $\{\theta_1,...,\theta_m\}$, where each $\theta_i$ maps $X_0$ to the
constant $a_i$. Moreover, $|\pi_{\chi(p')}(\sigma_{\theta_i} (r_{p'}))|=1$ holds. Therefore, we have $\mathit{bound}(\DB_2,\HD_2')=1$, and by
Theorem~\ref{thm:mainHybrid} we know that the counting problem can be efficiently solved over $\DB_2$ by using $\HD_2'$.  \hfill $\lhd$
\end{exa}

As the above example pointed out, to end up with better degree values, it could also be convenient to use higher values of the width. Indeed
the two parameters subtly interplay. Hence, we next face the natural problem of computing an optimal decomposition.

\begin{defi}\label{def:optimal}
Let $Q$ be a query, let $\DB$ be a database, and let $\mathcal{C}$ be a class of hypertree decompositions. Then, a hypertree decomposition
$\HD\in \mathcal{C}$ of $Q$ is $\DB$-\emph{optimal over $\mathcal{C}$} if there is no hypertree decomposition $\HD'\in\mathcal{C}$ of $Q$ such
that $\mathit{bound}(\DB,\HD') < \mathit{bound}(\DB,\HD)$.~\hfill~$\Box$
\end{defi}

Let $\kHD$ denote the class of all width-$k$ hypertree decompositions.
We first show a negative result for the computation of optimal decompositions over $\kHD$, due to a too liberal interplay between the structure
and the data in Definition~\ref{def:optimal}.

\begin{thm}\label{thm:NP}
Let $k\geq 8$ be any natural number. Computing $\DB$-optimal decompositions over $\mathcal{C}_{k}$ is $\rm NP$-hard.
\end{thm}
\begin{proof}
Recall that deciding whether there is a width-$k$ \emph{query decomposition}~\cite{chek-raja-97} is  $\rm NP$-hard, for every fixed $k\geq
4$~\cite{gott-etal-99}. Such decompositions can be identified with those hypertree decompositions $\HD=\tuple{T,\chi,\lambda}$ of $Q$ such that
for each vertex $p$, $\chi(p)=\vars(\lambda(p))$.

Observe that the problem is still $\NP$-hard if we look for the existence of {\em full width decompositions}, i.e., decompositions where every
vertex $p$ has precisely $|\lambda(p)|=k$ atoms, and if we consider queries where each atom has a distinguished variable (not occurring
elsewhere). This is easily seen. Given any query $Q$, build a query $\bar Q$ where each atom $q\in\atoms(Q)$ is replaced by $k\times
|\atoms(Q)|$ copies over fresh relation symbols and with the same set of variables plus a distinguished fresh variable for each of them. It can
be checked that $Q$ has a width-$k$ query decomposition if, and only if, $\bar Q$ has a width-$k$ query decomposition where moreover each
vertex $p$ is such that $|\lambda(p)|=k$.

\smallskip

We show a polynomial-time reduction from the (full) width-$k$ query decomposition problem to the problem of computing any optimal decomposition
over $\mathcal{C}_{2k}$. Let $Q$ be an instance of the full width-$k$ query decomposition problem enjoying the property discussed above, with
$\atoms(Q)=\{q_1,...,q_n\}$ and $\vars(Q)=\{Y_1,...,Y_z\}$.

Based on $Q$, we build in polynomial time a query $Q'$ and a database $\DB$ as follows. The set $\vars(Q')$ includes all variables in
$\vars(Q)$, plus a fresh variable $X_i$, for each $i\in\{1,...,n\}$. We define $\free(Q')=\{X_1,..,X_n\}$.
Moreover, we define $\atoms(Q')=\atoms(Q)\cup \{q_1',...,q_n'\}$, where for each $i\in\{1,...,n\}$, $\vars(q_i')=\vars(q_i)\cup\{X_i\}$.
Note that for each atom $q_i'$, with $i\in\{1,...,n\}$, there is a variable $S_i\in \vars(q_i')$ that occurs only in the corresponding unprimed
query atom $q_i$. Moreover, the variable $X_i$ occurs in $q_i'$ only.

The database $\DB$ is defined over the constants in the set $\{c_0,c_1,...,c_n\}$. For each $i\in\{0,..,n\}$, let $\theta_i$ be the
substitution mapping each variable in $\atoms(Q')$ to the constant $c_i$. Then, for each atom $q_j\in \atoms(Q)$, we define $q_j^\onDB$ as the
set $\{ \pi_{\vars(q_j)}(\{\theta_i\}) \mid i\in\{1,..,n\}\}$. So, $|q_j^\onDB|=n$. Moreover, for each $q_j'\in \atoms(Q')\setminus \atoms(Q)$,
we define
$$
\begin{array}{ll}
q_j'^{\onDB} = &\pi_{X_j}(\{\theta_0\})\times \{ \pi_{\vars(q_j)}(\{\theta_i\}) \mid i\in\{1,..,n\}\setminus\{j\}\} \cup \\
               &\pi_{X_j}(\{\theta_j\})\times r_{\mbox{-}j},
\end{array}
$$
where $\times$ denotes the standard Cartesian product, and $r_{\mbox{-}j}$ is the set of all substitutions for the variables in $\vars(q_j)$
where one variable is mapped to $c_j$ and all other variables are mapped to the same constant in $\{c_1,...,c_n\}$ (possibly coinciding with
$c_j$). So, $|r_{\mbox{-}j}|=|\vars(q_j)|\times n$ holds.

In order to conclude, we claim that a hypertree decomposition $\HD'\in\mathcal{C}_{2\times k}$ of $Q'$ exists with $\mathit{bound}(\DB,\HD)\leq
n-k$ if, and only if, $Q$ admits a full width-$k$ query decomposition.

Let $\HD'=\tuple{T,\chi',\lambda'}$ be a width-$(2\times k)$ hypertree decomposition of $Q'$, and assume that $\mathit{bound}(\DB,\HD')\leq
n-k$. Because of the definition of $q_j'^{\onDB}$, for any atom of the form $q_j'$ with $j\in\{1,...,n\}$, the following properties hold:

   \begin{enumerate}
    \item the number of substitutions in $q_j'^{\onDB}$ assigning the value $c_0$ to the free variable $X_j$ are $n-1$ (only the value
        $c_j$ is missing in the assignment to the remaining variables);
    \item the substitutions assigning the value $c_j$ to the free variable $X_j$ are designed in a way that they find a matching
        substitution over existential variables in every other relation, with the only exception of those relations for atoms having
        precisely the same set of existential variables as $q_j'$. However, by construction, only $q_j$ has precisely the same existential
        variables (every other atom differs at least on its distinguished variable).
   \end{enumerate}

It follows that, in order to deal with the above sources of degree unboundness, and get the desired degree-value below $n-k$, at each vertex $p$ in
$T$, it must be the case that: $|\{q_1',...,q_n'\} \cap \lambda'(p)|\geq k$; for each $q_i'\in \lambda'(p)$, $q_i$ belongs to $\lambda'(p)$;
and $\chi'(p)\supseteq (\vars(\lambda'(p))\setminus\free(Q))$. Given the width of $\HD'$, this entails that $\lambda'(p)=\{q_{\alpha_1},
q_{\alpha_1}',...,q_{\alpha_k}, q_{\alpha_k}' \}$, with $|\lambda'(p)|=2\times k$.
Therefore, based on $\HD'$ we can immediately build a novel hypertree $\HD=\tuple{T,\chi,\lambda}$ such that $\lambda(p)=\lambda'(p)\cap
\{q_1,...,q_n\}$ and $\chi(p)=\vars(\lambda(p))$, which is a full width-$k$ query decomposition of the original query $Q$.

On the other hand, if $\HD=\tuple{T,\chi,\lambda}$ is a full width-$k$ query decomposition of the query $Q$, then we build a query
decomposition $\HD'=\tuple{T,\chi',\lambda'}$ of $Q'$ such that $\lambda'(p)=\lambda(p) \cup\{ q_j' \mid q_j\in \lambda(p)\}$. In fact, the
width of $\HD'$ is $2\times k$, and $\mathit{bound}(\DB,\HD)\leq n-k$.
\end{proof}

Intuitively, the source for the above intractability is that, according to its definition, a hypertree decomposition might mix together in some
vertex variables and atoms that might instead be evaluated independently. Usually, this is not a main issue. Indeed, \emph{normal forms} have
been defined~\cite{SGL04,GS08}, which typically correspond to more efficient decompositions, and it has been shown that focusing on normal
forms is not restrictive, as any given hypertree decomposition can be transformed into a normal form one by preserving its width. Here, things
are different as, by mixing in some vertex variables and atoms that might be evaluated independently, we can in principle end up with a better
degree value. Our intuition is confirmed by the result below, evidencing that over normal form decompositions the picture is completely
different.
Let $\kNFHD$ be the class of all width-$k$ hypertree decompositions in normal form.

\begin{thm}
Computing $\DB$-optimal decompositions over $\kNFHD$ is feasible in polynomial time.
\end{thm}
\begin{proof}
Let $Q$ be a query and let $\DB$ a database for it. Deciding whether $Q$ has a hypertree decomposition in  $\kNFHD$ is feasible in polynomial
time~\cite{gott-etal-99}. Therefore, let us focus on the case where the goal is precisely to compute any $\DB$-optimal decomposition in normal
form therein.

Let $\mathcal{N}$ be the set of all possible vertices of any width-$k$ hypertree decomposition $\tuple{T,\chi,\lambda}$ for the query $Q$.

Consider the function $v_{\onDB}:\mathcal{N}\mapsto \mathbb{R^+}$ such that, $\forall p\in \mathcal{N}$,
$$v_{\onDB}(p)=\max_{\theta \in \pi_{\free(Q)}(r_p) } {(w+1)}^{|\sigma_\theta(r_p)|},$$
where $r_p=\pi_{\chi(p)}(\bigwedge_{q\in \lambda(p)}q)^\onDB$ and $w=|\atoms(Q)|$.

Note that $|\sigma_\theta(r_p)|\leq m$ holds for each $\theta \in \pi_{\free(Q)}(r_p)$, so that the size of $v_{\onDB}(p)$ is polynomial
w.r.t.~$m$.
Moreover, consider the aggregate function $F_{Q,\DB}$ associating with each hypertree $\HD=\tuple{T,\chi,\lambda}$ of $Q$ in $\kNFHD$ the real
number $F_{Q,\DB}=\sum_{p\in \vertices(T)}v_{\onDB}(p)$.

\smallskip

Consider now the hypertree $\HD^*\in\kNFHD$ of $Q$ such that $F_{Q,\DB}(\HD^*)<F_{Q,\DB}(\HD')$, for each hypertree $\HD'\in \kNFHD$ of $Q$.
We claim that $\HD^*$ is $\DB$-optimal over $\kNFHD$. Indeed, let $h=\mathit{bound}(\DB,\HD^*)$, and observe that $F_{Q,\DB}(\HD^*)\geq (w+1)^h$.
Assume,  for the sake for contradiction, that $\HD^*$ is not $\DB$-optimal over $\kNFHD$ and let $\HD'\in \kNFHD$ be another hypertree
decomposition of $Q$ such that $\mathit{bound}(\DB,\HD')=h'\leq h-1$. Since $\HD'$ is in normal form, then it has at most $w$ vertices~(in
fact, the number of its vertices is bounded by the minimum between the number of atoms and the number of variables occurring in
$Q$)\cite{SGL04,GS08}. Let $n_{h'}$ be the number of vertices $p$ of $\HD'$ such that $v_{\onDB}(p)=h'$, and observe that:
$$
\begin{array}{l}
F_{Q,\DB}(\HD')\leq n_{h'}\times (w+1)^{h'}+(w-n_{h'})\times (w+1)^{h'-1}.
\end{array}
$$

Hence, $F_{Q,\DB}(\HD')\leq w\times (w+1)^{h'}< (w+1)^{h'+1}\leq (w+1)^h\leq F_{Q,\DB}(\HD^*)$ holds, which is impossible.

To conclude, we point out that the function $F_{Q,\DB}$ fits the framework of the {weighted} hypertree decompositions defined in \cite{SGL04}.
There, it is shown that the hypertree $\HD^*$ can be computed in polynomial time, for any fixed natural number $k$, so that our result follows from the above claim.
\end{proof}

\end{appendix}